\documentclass{article} 
\usepackage{iclr2024_conference,times}

\usepackage{amsmath,amsfonts,bm}

\def\eqref#1{equation~\ref{#1}}

\def\1{\bm{1}}

\DeclareMathAlphabet{\mathsfit}{\encodingdefault}{\sfdefault}{m}{sl}
\SetMathAlphabet{\mathsfit}{bold}{\encodingdefault}{\sfdefault}{bx}{n}

\iclrfinalcopy
\usepackage{xspace}
\usepackage[colorlinks, linkcolor=red, anchorcolor=red, citecolor=cyan, breaklinks]{hyperref}
\usepackage{longtable}
\usepackage{array}
\usepackage{url}
\usepackage{makecell}
\usepackage{markdown}
\usepackage{arydshln}
\usepackage{soul}
\usepackage[normalem]{ulem}
\usepackage{mdframed}
\usepackage{xcolor}
\usepackage{ltablex}
\usepackage[utf8]{inputenc} 
\usepackage[T1]{fontenc}    
\usepackage{booktabs}       
\usepackage{amsfonts}       
\usepackage{nicefrac}       
\usepackage{microtype}      
\usepackage{xcolor}         
\usepackage{microtype}
\usepackage{graphicx}
\usepackage{tabularx}
\usepackage{subcaption}
     
\usepackage{booktabs}
\usepackage{hyperref}
\usepackage{color}
\usepackage{xcolor}
\usepackage[breakable, skins, xparse]{tcolorbox}

\usepackage{amsmath}
\usepackage{amssymb}
\usepackage{mathtools}
\usepackage{amsthm}
\usepackage[capitalize,noabbrev]{cleveref}
\usepackage{algorithm}
\usepackage{algorithmic}

\usepackage[textsize=tiny]{todonotes}
\usepackage{multirow}
\usepackage{float}
\usepackage{stfloats}
\usepackage{times}
\usepackage{epsfig}
\usepackage{comment}
\usepackage{wrapfig}
\usepackage{enumitem}
\usepackage{subcaption}
\usepackage{subfiles}

\usepackage{color}
\usepackage{colortbl}
\setlist[itemize]{leftmargin=2em}

\usepackage{wrapfig}

\usepackage{amssymb}
\usepackage{amsbsy}
\usepackage{multirow}

\usepackage{bbding}
\usepackage{color}
\usepackage{fancyhdr}

\renewcommand{\algorithmicrequire}{\textbf{Input:}}
\renewcommand{\algorithmicensure}{\textbf{Output:}}
\newcommand{\model}{AI-Trader}

\newcommand{\uparrowred}{\textcolor{red}{$\uparrow$}}
\newcommand{\downarrowgreen}{\textcolor{green!60!black}{$\downarrow$}}

\usepackage{wrapfig}

\title{AI-Trader: Benchmarking Autonomous Agents in Real-Time Financial Markets}

\author{Tianyu Fan, Yuhao Yang, Yangqin Jiang, Yifei Zhang, Yuxuan Chen, Chao Huang\thanks{Corresponding Author: Chao Huang}\\
University of Hong Kong\\
{\tt $\{$tianyufan$\}$@connect.hku.hk}; \\~\\
\textbf{Website:} \href{https://ai4trade.ai/}{\xspace\texttt{https://ai4trade.ai/}}\\
}

\begin{document}

\maketitle

\begin{abstract}

Large Language Models (LLMs) have demonstrated remarkable potential as autonomous agents, approaching human-expert performance through advanced reasoning and tool orchestration. However, decision-making in fully dynamic and live environments remains highly challenging, requiring real-time information integration and adaptive responses. While existing efforts have explored live evaluation mechanisms in structured tasks, a critical gap remains in systematic benchmarking for real-world applications, particularly in finance where stringent requirements exist for live strategic responsiveness. To address this gap, we introduce \model, the first fully-automated, live, and data-uncontaminated evaluation benchmark for LLM agents in financial decision-making. \model\ spans three major financial markets: U.S. stocks, A-shares, and cryptocurrencies, with multiple trading granularities to simulate live financial environments. Our benchmark implements a revolutionary fully autonomous minimal information paradigm where agents receive only essential context and must independently search, verify, and synthesize live market information without human intervention. We evaluate six mainstream LLMs across three markets and multiple trading frequencies. Our analysis reveals striking findings: general intelligence does not automatically translate to effective trading capability, with most agents exhibiting poor returns and weak risk management. We demonstrate that risk control capability determines cross-market robustness, and that AI trading strategies achieve excess returns more readily in highly liquid markets than policy-driven environments. These findings expose critical limitations in current autonomous agents and provide clear directions for future improvements. The code and evaluation data are open-sourced to foster community research: \url{https://github.com/HKUDS/AI-Trader}.
\end{abstract}

\section{Introduction} \label{intro}

Large Language Models have catalyzed a transformative shift toward autonomous agents, endowing these systems with sophisticated capabilities in complex reasoning, dynamic tool orchestration, and strategic long-horizon planning~\citep{gao2024agentscope,tang2025autoagent,zhang2025agentorchestrahierarchicalmultiagentframework}. However, this remarkable progress has exposed a critical evaluation gap, particularly in dynamic and real-time scenarios that represent some of the most demanding real-world applications for autonomous systems. Traditional static benchmarks concentrate on question answering~\citep{yang2018hotpotqa,joshi2017triviaqa,trivedi2022musique}, code completion~\citep{ding2023crosscodeeval,chen2024ppm,xie2025core,zhang2025artifactsbench}, and single-turn instruction following~\citep{bai2024longwriter,jiang2024followbench}, operating within constrained, deterministic environments with limited real-world complexity.

Financial markets present a stark contrast to these static evaluation environments. Unlike traditional benchmarks with fixed datasets, financial markets are inherently dynamic, continuous systems characterized by extreme volatility and unpredictable shifts. This dynamic complexity makes financial markets ideal testbeds for evaluating fundamental agent abilities including planning, information retrieval, numerical reasoning, and real-time decision-making under uncertainty. Moreover, performance in financial domains admits a simple yet objective evaluation metric: investment returns. These dynamic scenarios present fundamentally different challenges that require agents to navigate uncertainty, adapt to rapidly evolving conditions, and execute consequential decisions under stringent time constraints. Conventional static evaluation approaches fundamentally fail to assess agents' performance in such high-stakes financial environments that demand instantaneous market analysis, sophisticated risk assessment, and time-critical strategic decision-making under profound uncertainty.

Recent research has explored more sophisticated evaluation paradigms by integrating tool-use capabilities to test autonomous agents in dynamic scenarios~\citep{wu2025webwalker,xie2024osworld,jia2025osworld,wei2025browsecomp,chen2025xbench}. However, these frameworks fundamentally struggle to handle the dynamic and real-time nature of financial domains due to critical limitations. Current approaches lack sufficient automation and are heavily dependent on fixed prompts and predefined workflows, rendering them ineffective in dynamic real-time financial scenarios that require continuous autonomous adaptation. In addition, the absence of genuine time constraints, market volatility, and economic consequences creates a significant disconnect between benchmark performance and actual real-world capability. This evaluation gap makes current approaches inadequate for assessing agents' true potential. Complex financial scenarios demand continuous adaptability and robust decision-making. Agents must perform under rapidly evolving market conditions.

To address these fundamental limitations and push the boundaries of agent capabilities, we design a benchmark that involves \textbf{long-horizon tool use}, a \textbf{continuously evolving environment}, and tasks that are \textbf{challenging to execute yet straightforward to evaluate}~\citep{zeng2025futurex,jain2024livecodebench}. We propose \textbf{AI-Trader}, the first fully autonomous, live, data-uncontaminated evaluation environment spanning three major markets: U.S. stocks, A-shares, and cryptocurrencies. \model\ operates in real-time market conditions with strict temporal filtering and carefully designed tool interfaces. This enables realistic assessment of agent capabilities under actual market volatility and uncertainty.

\model\ introduces a \textbf{fully autonomous minimal information paradigm} that operates with complete automation and unprecedented independence. Agents receive only essential context: available tools, current portfolio holdings, and real-time market prices. No human intervention or pre-packaged information is provided throughout the process. Agents must autonomously search live internet and financial markets to gather relevant data, synthesize comprehensive insights, and generate well-informed trading decisions through their tool-use pipelines. This revolutionary design completely eliminates human input and delegates full execution authority to the agent. It forces rigorous autonomous capability demonstration in information acquisition, complex synthesis, and strategic decision-making under time-critical constraints in real-time financial environments.

Our comprehensive evaluation across six mainstream LLMs reveals striking performance disparities. These critical limitations are invisible in static benchmarks. We demonstrate that general intelligence does not automatically translate to effective trading capability. Most agents exhibit poor returns and weak risk management in fully autonomous operations. Our analysis uncovers that AI trading strategies achieve excess returns more readily in highly liquid markets (U.S.) compared to policy-driven environments (A-shares). Model generalization capabilities exhibit significant cross-market limitations when operating without human guidance. Key contributions of this work are as follows:

\begin{itemize}[leftmargin = *]
    \item \textbf{First Fully Autonomous Live Multi-Market Evaluation Platform}: We introduce the first financial evaluation benchmark that is completely agent-operated, real-time, and data-uncontaminated. Our platform spans U.S. equities, A-shares, and cryptocurrencies. It incorporates rigorous time consistency and information isolation mechanisms. This fundamentally breaks away from traditional static evaluation paradigms and establishes a new benchmark for autonomous agent assessment.
    \item \textbf{End-to-End Autonomous Assessment Framework}: Agents must independently search, verify, and synthesize live market information. They use only minimal context: current holdings, real-time prices, and available tools. This enables comprehensive evaluation of long-horizon tool utilization and adaptive reasoning. Decision-making capabilities are tested under genuine market constraints. 
    \item \textbf{Systematic Cross-Market Agent Capability Analysis}: We evaluate six mainstream LLMs across three distinct markets and multiple trading frequencies. We systematically analyze performance in profitability, decision-making processes, and tool usage patterns. Our findings reveal critical limitations in autonomous execution, risk management, and cross-market generalization. These limitations are invisible in static benchmarks, highlighting the necessity for live evaluation.

\end{itemize}

\section{Related Works}

$\bullet$ \textbf{LLM Agents for Autonomous Decision-Making.}
Large language models have undergone a transformative evolution from powerful text-completion systems into autonomous agents. These systems are now capable of sophisticated reasoning, strategic planning, and dynamic interaction with external environments~\cite{Anthropic2025a,Anthropic2025b,qwen2,tang2025autoagent,xie2024osworld,openai2023gpt4}. This emergent agent capability represents the next frontier in LLM development. It directly bridges linguistic competence with tangible productivity and economic value in real-world applications~\citep{patwardhan2025gdpval,arora2025healthbench}. The core strength of LLM agents lies in their capacity for long-horizon decision-making under uncertainty. They can maintain coherent strategies across extended time periods. Simultaneously, they adapt to evolving circumstances and complex environmental changes. \model\ leverages these advanced agent capabilities to enable autonomous decision-making in dynamic financial markets without human intervention.

$\bullet$ \textbf{Tool Use and Environmental Interaction.}
To realize autonomous capabilities, LLM agents must interact with their environments through sophisticated tool use. This enables them to invoke external functions, APIs, and services to acquire real-time information, perform computations, and execute actions~\citep{li2025,Mohammadi2025,Lu2025}. Tool use is fundamental to sustaining the perception–decision–action loop that defines autonomous agency~\citep{Yehudai2025,Raseed2025}. Recent advances have accelerated this development significantly. Standardized protocols like the Model Context Protocol (MCP)~\citep{MCP2025} provide unified interfaces for LLMs to access and orchestrate external tools efficiently. This broadens the scope and interoperability of LLM-based agents across diverse domains. Our framework advances these paradigms by implementing a fully-automated minimal information approach specifically designed for financial markets. We provide agents with only essential trading tools while requiring them to independently search, verify, and synthesize live market information with zero human supervision or guidance.

$\bullet$ \textbf{From Static to Live Agent Evaluation}.
To systematically evaluate emerging agent capabilities in dynamic environments, researchers have developed sophisticated benchmarks that simulate complex real-world decision-making processes. These live-environment benchmarks fall into two distinct categories based on their design philosophy and evaluation scope. The first type combines static question banks with dynamic execution environments. Examples include SWE-Bench~\citep{jimenez2023swe} for software engineering tasks, GAIA~\citep{mialon2023gaia} and BrowseComp~\citep{wei2025browsecomp} for web-based information retrieval, and XBench~\citep{chen2025xbench} and Tau2Bench~\citep{yao2024tau} for workflow automation. The second type features complete dynamicity in both task sets and execution environments. LiveCodeBench~\citep{jain2024livecodebench} and FutureX~\citep{zeng2025futurex} continuously update their task collections through automated pipelines. This better reflects the evolving nature of real-world scenarios where agents must adapt to time-evolving conditions.

Financial decision-making presents unique challenges that make it an ideal testbed for agent evaluation. It inherently involves dynamic market responses, real-time information integration, and sequential action selection under high stakes. Existing financial benchmarks span the static–dynamic spectrum. Static approaches like FinSearchComp~\citep{hu2025finsearchcomp} and FinBench~\citep{lu2025bizfinbench} assess factual knowledge. Dynamic approaches including InvestorBench~\citep{li-etal-2025-investorbench}, Finance Agent Benchmark~\citep{bigeard2025finance}, LiveTradeBench~\citep{yu2025livetradebench}, and StockBench~\citep{chen2025stockbench} attempt to simulate realistic trading conditions. However, these approaches still lack the genuine autonomy and real-time constraints that characterize actual financial environments.

\section{The Framework of \model}

\begin{figure}
    \centering
\includegraphics[width=0.9\linewidth]{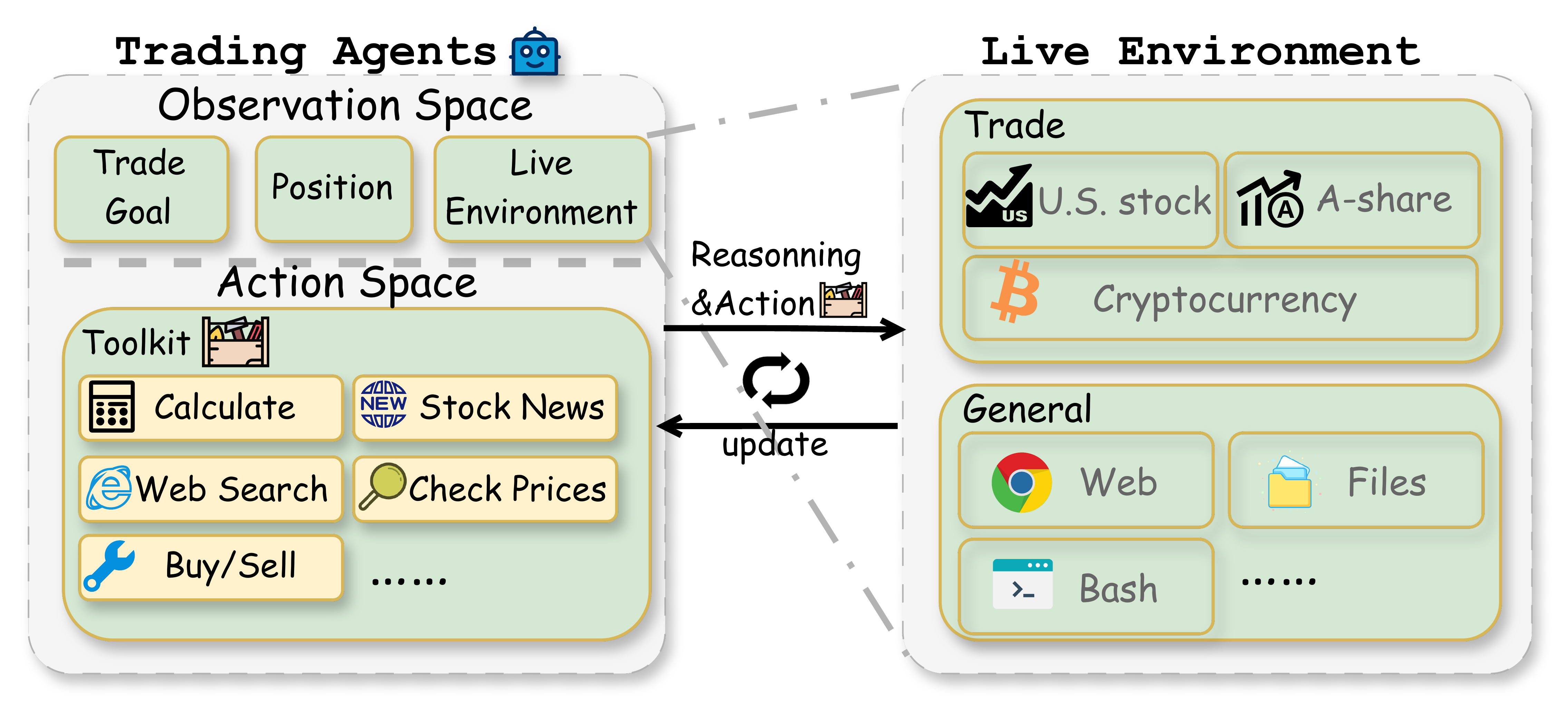}
\caption{The architecture of \model. In \model, all information must be acquired through tools, ensuring that the decisions and actions generated by the agent can be observed under fully autonomous behavior. We equip the agent with three mainstream trading environments: U.S. stocks, A-share stocks, and cryptocurrencies. Additionally, we provide five fundamental tools that enable the agent not only to interact with these trading environments but also to perform computations via Bash, read local files to access stock data, and browse the web for general information.}

    \label{fig:mainmodel}
\end{figure}

 The \model~benchmark aims to provide a \textbf{high-fidelity}, \textbf{data-uncontaminated}, and \textbf{fully autonomous} experimental environment for evaluating agent decision-making capabilities in highly dynamic financial markets. Our design philosophy centers on simulating the complete workflow of professional financial analysts. This includes real-time market research, strategic reasoning, and autonomous trading execution. We eliminate all forms of human intervention and delegate full execution authority to the agent at every step. This approach enables systematic, end-to-end assessment of agent capabilities under genuine market conditions. Unlike traditional benchmarks that rely on historical data or simulated environments, our framework operates in live markets with real economic consequences. Agents must navigate authentic time pressures, market volatility, and information uncertainty that characterize actual financial decision-making. The \model\ framework consists of two fully decoupled components: the Live Trading Environment (Sec.\ref{sec_liveenv}) and the Trading Agents (Sec.~\ref{sec_tradeagent}). This separation ensures evaluation integrity while providing agents with the flexibility to develop their own trading strategies and decision-making processes.

\subsection{Live Trading Environment}
\label{sec_liveenv}
The \model\ benchmark operates across three distinct global financial markets to systematically assess agent generalization and adaptability under diverse conditions. We select the U.S. stock market, China's A-share market, and the cryptocurrency market as they represent fundamentally different regulatory frameworks, investor compositions, and market dynamics. Additionally, we support both hourly and daily trading frequencies to capture comprehensive market behaviors and test agent responsiveness across multiple time horizons.

\subsubsection{U.S. Stock Market}
The U.S. stock market serves as our primary testing environment due to its institutional maturity and global influence. It combines high liquidity, transparent information flow, and strong sensitivity to technological innovation and macroeconomic factors. This creates an ideal benchmark for evaluating agent decision-making capabilities under sophisticated market conditions.

$\bullet$ \textbf{Market Characteristics.}
The U.S. equity market represents the world's most liquid and institutionally developed trading environment. It features transparent information disclosure and sophisticated institutional participation. The market serves as a global benchmark for equity valuation and risk assessment. Market movements are primarily driven by Federal Reserve policy, corporate earnings, technological innovation, and macroeconomic indicators. This creates a dynamic environment that balances long-term growth trends with periodic volatility.

$\bullet$ \textbf{Portfolio Construction.}
To construct our U.S. portfolio, we select all constituents of the Nasdaq-100 Index\footnote{\url{https://www.nasdaq.com/market-activity/index/ndx}}. This index comprises the 100 largest non-financial companies listed on the Nasdaq Stock Market. It represents the forefront of technological innovation across multiple sectors. These include information technology, semiconductors, biotechnology, internet services, and consumer discretionary. Key constituents include Apple, Microsoft, NVIDIA, Amazon, Alphabet, Tesla, and Meta. These are dominant enterprises that exhibit liquidity, R\&D intensity, and growth potential.

$\bullet$ \textbf{Evaluation Advantages.}
This selection provides several advantages for agent evaluation. First, these companies demonstrate high market sensitivity to technological developments and macroeconomic shifts. This creates rich information environments for testing agent adaptability. Second, the portfolio focuses on growth-oriented, innovation-driven companies. This ensures exposure to dynamic market conditions that require sophisticated analytical capabilities. Additionally, we include a cash asset with zero return to enable risk-free capital allocation strategies. This allows agents to demonstrate portfolio management skills across the risk spectrum.

\subsubsection{A-share Market}
The A-share market provides a contrasting environment to test agent adaptability under different regulatory and structural conditions. Its unique combination of market dynamics, diverse investor composition, and evolving frameworks creates distinct challenges for systematic trading strategies.

$\bullet$ \textbf{Market Characteristics.}
The A-share market serves as China's core equity trading platform, featuring distinct regulatory frameworks and diverse investor composition. Market mechanisms continue to improve with increasingly robust disclosure and regulatory frameworks. A-share prices are driven by macroeconomic expectations, sector dynamics, and market sentiment. This results in episodic volatility coexisting with structural opportunities. Effective portfolio management in this environment requires balanced integration of macroeconomic analysis, sector rotation identification, and adaptability to non-stationary market behaviors.

$\bullet$ \textbf{Portfolio Construction.}
To construct our A-share portfolio, we select all 50 constituent stocks of the SSE-50 Index\footnote{\url{https://english.sse.com.cn/indices/indices/list/indexmethods/c/000016_000016hboeken_EN.pdf}}. This index comprises the largest and most liquid blue-chip companies listed on the Shanghai Stock Exchange. It spans key economic sectors including finance, consumer goods, industrials, information technology, energy, and healthcare. The index collectively reflects the performance of China's high-quality listed enterprises across multiple industries. Constituents include industry leaders such as Ping An Insurance, Kweichow Moutai, China Merchants Bank, Industrial and Commercial Bank of China, and Contemporary Amperex Technology.

$\bullet$ \textbf{Evaluation Advantages.}
This selection offers strong market representativeness and liquidity for agent evaluation. The market-sensitive nature of these stocks tests agent capabilities in interpreting economic signals and macroeconomic trends. The diverse sector composition allows assessment of cross-industry analysis skills. Additionally, the market's unique dynamics provide opportunities to evaluate agent performance under different investor behavior patterns compared to Western markets.

\subsubsection{Cryptocurrency Market}
The cryptocurrency market operates as a continuous, globally distributed ecosystem with extreme volatility and 24/7 trading. This environment tests agent capabilities under constant market pressure, rapid information flow, and heightened uncertainty that characterizes digital asset trading.

$\bullet$ \textbf{Market Characteristics.}
The cryptocurrency market represents a decentralized digital asset ecosystem that operates continuously without traditional market hours. It features extreme volatility, technology-driven dynamics, and exceptional sensitivity to regulatory developments and market sentiment. Although blockchain technologies and real-world use cases continue to evolve, price movements remain heavily influenced by technological breakthroughs, regulatory announcements, institutional adoption pace, and macroeconomic conditions. This creates a distinct market profile where high volatility coexists with high potential returns and pronounced cyclical patterns.

$\bullet$ \textbf{Portfolio Construction.}
To construct our cryptocurrency portfolio, we select 10 major trading pairs tracked by the Bitwise Index, all quoted in USDT: BTC, ETH, XRP, SOL, ADA, SUI, LINK, AVAX, LTC, and DOT. This encompasses diverse blockchain functionalities across the ecosystem. It includes store-of-value assets (BTC), smart contract platforms (ETH, SOL, ADA, DOT, SUI), cross-border payment solutions (XRP), oracle protocols (LINK), and established currencies (LTC). These assets represent leading market capitalizations, liquidity, and established technological utility.

$\bullet$ \textbf{Evaluation Advantages.}
This portfolio provides exposure to key innovation domains including decentralized finance, Web3 infrastructure, Layer 1 and Layer 2 scaling solutions, and cross-chain interoperability. The 24/7 trading environment tests agent responsiveness under continuous market conditions. High volatility and rapid information flow challenge agents to process news, technical developments, and sentiment shifts in real-time. Additionally, the technology-focused nature of crypto markets evaluates agent capabilities in interpreting innovations and market implications.

\subsection{Trading Agents}
\label{sec_tradeagent}

The design of \model~aims to serve as an easy-to-use testing environment for agents. Accordingly, \model~is designed to be easily integrable and readily modifiable. Its agent architecture is modular, separating the core system, MCP toolchain, data system, and prompt engineering components, and connecting them via standardized interfaces. This enables different agents to autonomously perform market data queries, trade execution, information retrieval, and strategy reasoning under a unified set of rules. To facilitate convenient third-party testing, \model~supports a registration mechanism for custom agents and MCP tools, allowing developers to effortlessly inject new strategies or extend functionalities—thus creating a plug-and-play, extensible testing platform.

\subsubsection{Design of Base Trading Agents}

As the initiator of decisions, the agent must transform raw observations obtained through tools into explicit, executable trading actions via structured reasoning. Immediately after executing a decision, the agent must perceive environmental feedback—including changes in asset prices, portfolio positions, and market news updates—thereby forming a new round of observations. This process establishes a continuous, closed-loop, adaptive Observe-Reason-Act cycle.
For \model~, its core design principles are: all information must be acquired through tools, all decisions must be based on autonomous reasoning, and all actions must be executable under real-world constraints.

\textbf{Observation Space.}
For a trading agent $f$, its initial perception of the market state consists primarily of two components:  
1. The current price of assets $ \mathbf{p} = [p_1, p_2, \dots, p_n] $, which reflects real-time market conditions;  
2. The agent’s own current holding status $ \mathbf{s} = [s_1, s_2, \dots, s_n]$, which characterizes its existing positions.  
This information forms the initial observation $ o_0 $. On top of this, the agent can dynamically acquire additional information through tool invocations—for example, detailed data $ \pi_i $ for specific stocks (such as fundamental or technical indicators), as well as broader general information $ i $ (such as market news or macroeconomic indicators). All of this information is integrated into the agent’s complete perception at the current time step, denoted as $ o_t = f(\mathbf{p}, \mathbf{s}, \{\pi_i\}, i) $, which serves as the primary basis for its subsequent reasoning and trading decisions.

\textbf{Reasoning Process.}
After obtaining the observation $ O_t $, the agent enters a fully autonomous reasoning phase that does not rely on any external intervention or pre-defined logic. This reasoning follows the ReAct~\citep{yao2022react} paradigm, emphasizing ``think first, act later''.
In \model, the reasoning process can yield one of two outcomes: either invoking another tool to gather additional observations or directly making a trading decision. Throughout this process, the agent generates intermediate reasoning in natural language, such as ``A company’s Q2 earnings significantly beat expectations, and its P/E ratio is below the industry average, indicating it is a compelling buy.''
These natural language reasoning traces are fully recorded, ensuring the entire decision-making process is observable, auditable, and reproducible. This design not only enhances the agent’s adaptability in complex market environments but also enables researchers to further analyze the agent’s behavior in financial decision-making.

\textbf{Action Space.}
The agent's action generation is strictly based on the aforementioned observations $o_t$ and autonomous reasoning $r_t$, ensuring its behavior is fully autonomous, resource-constrained, and compliant with market regulations. For each tradable asset, the agent may execute only one of three discrete actions $a_t$ : (1) increase position (Buy), (2) decrease position (Sell), or (3) hold. This design avoids an overly complex continuous action space and focuses on interpretable, executable fundamental decision units.
In the \model, if a proposed action would cause the total capital requirement to exceed available liquidity, the system rejects execution and triggers the agent’s self-correction mechanism—requiring it to re-invoke tools, re-reason, and generate a new decision that satisfies both liquidity and compliance constraints.
Ultimately, the action is generated by the policy function $ a_t = f(o_t, r_t) $, where $ o_t $ denotes the observation and $ r_t $ denotes the reasoning. This entire action pipeline forms a closed loop—from information acquisition and autonomous reasoning to decision execution—realistically simulating the complete decision-making chain of an autonomous trading agent in a multi-asset market.

\subsubsection{Tool chain for Agents}
In this section, we present the minimal set of tools designed within the \model~framework to support the core functionalities of a Trading Agent. This toolkit is intended to provide the essential capabilities required to perform fundamental trading tasks, including market data retrieval, order execution, and information seeking. During the design process, we placed particular emphasis on modularity and extensibility of the architecture, enabling seamless adaptation to future requirements across diverse asset classes (e.g., equities, cryptocurrencies) and trading frequencies (e.g., daily, hourly, or real-time trading). To this end, all tools are uniformly built atop the Model Context Protocol~\citep{MCP2025}. Through this design, \model~maintains a lightweight core while flexibly supporting a wide range of dynamic and evolving trading scenarios.

\textbf{Check Price.} The Check Price tool is designed to provide AI with accurate and timely stock price data. It can query the historical or intraday open, high, low, close prices, and trading volume of a specified stock based on the current simulated time. The tool automatically recognizes stock tickers across different markets and returns data that complies with each market’s trading rules, ensuring AI has reliable market information for decision-making.

\textbf{Search.} The Search tool retrieves publicly available information related to markets, companies, or macroeconomic factors. Driven by keyword queries, it performs real-time searches and returns news articles, official announcements, analyst reports, and other relevant content—strictly limited to information available up to the current simulated time—to prevent data leakage from the future and support AI in fundamental or event-driven analysis.

\textbf{News.} The News tool focuses on delivering structured financial news along with associated sentiment signals to help AI quickly grasp significant market-moving events. Its output includes publication time, related securities, and news summaries, specifically optimized for financial decision-making scenarios, thereby complementing general-purpose search with high-quality content from specialized financial media.

\textbf{Math.} The Math tool is designed to support AI in performing essential numerical calculations during the trading process.

\textbf{Trade.} The Trade tool executes actual buy and sell orders. Based on the agent’s decisions, it issues trading instructions while adhering to market-specific rules (e.g., A-share lot-size requirements of 100-share increments). It also updates portfolio holdings and cash balances in real time, ensuring the entire trading process is compliant, accurate, and fully auditable.

\section{Experimental Evaluation}
In this section, we present live trading results for U.S. equities and A-shares from October 1, 2025, to November 7, 2025, along with a detailed evaluation and analysis of the cryptocurrency market from November 1 to November 14, 2025. We employ a daily-frequency strategy for trading A-shares and cryptocurrencies, while monitoring and recording U.S. equity trades using an hourly-frequency strategy. This multi-market, multi-frequency design enhances the diversity and real-world representativeness of our experiments, enabling a more comprehensive assessment of the trading agent’s adaptability and robustness across different asset classes and temporal granularities. Furthermore, Sec.~\ref{exp_llm} and Sec.~\ref{exp_metric} describe the evaluation setup, including the model backbone and performance metrics used; Sec.~\ref{exp_preformance} reports the main results; Sec.~\ref{exp:casestudy} provides in-depth analysis and discussion through case study. For more detailed information, please refer to Appendix.~\ref{appdenix:prompt} and \ref{appendix:exp}.

\subsection{LLM backbones}
\label{exp_llm}
The \model~employs six mainstream models as baselines: DeepSeek-v3.1~\citep{liu2024deepseek}, MiniMax-M2~\citep{minimaxm2}, Claude-3.7-sonnet~\citep{claude3-7}
, GPT-5~\citep{gpt5}, Qwen3-max~\citep{yang2025qwen3}, and Gemini-2.5-flash~\citep{comanici2025gemini}. For all models, we encapsulate them into identically structured agents—each endowed with the exact same trading objectives and tools. This setup ensures a fair and objective comparison of the models' performance under identical task conditions, eliminating confounding factors arising from differences in agent architecture, tool invocation methods, or goal specifications. Consequently, any observed performance discrepancies can be confidently attributed to the models' intrinsic reasoning capabilities, decision-making strategies, and comprehension of financial scenarios.

\subsection{Metrics}
\label{exp_metric}

\begin{itemize}[leftmargin = *]
\item \textbf{Cumulative Return (CR).}  
This metric captures the total percentage gain or loss of a strategy over the entire evaluation period. It is computed as the compounded product of periodic returns minus one:  
$ \text{CR} = \left( \prod_{t=1}^{T} (1 + r_t) \right) - 1 $,  
where $ r_t $ denotes the return at time $ t $, and $ T $ is the total number of observation periods.

\item \textbf{Sortino Ratio (SR).}  
The Sortino Ratio evaluates risk-adjusted performance by penalizing only downside volatility relative to a specified target return (commonly zero). It is defined as the excess average return over the target return, divided by the downside deviation:  
$ \text{SR} = \frac{\bar{r} - r_{\text{target}}}{\sigma_d} $, where $ \sigma_d = \sqrt{ \frac{1}{T} \sum_{t=1}^{T} \bigl( \min(r_t - r_{\text{target}},\, 0) \bigr)^2 } $.  
Here, $ \bar{r} $ is the average return, and $ \sigma_d $ represents the downside deviation.

\item \textbf{Volatility (Vol).}  
Volatility measures the total risk of a strategy and is typically reported as the annualized standard deviation of returns. The sample standard deviation of returns is given by:  
$ \text{Vol} = \sqrt{ \frac{1}{T-1} \sum_{t=1}^{T} (r_t - \bar{r})^2 }$.

\item \textbf{Maximum Drawdown (MDD).}  
Maximum Drawdown quantifies the largest peak-to-trough decline in the cumulative value of a strategy during the evaluation period. Letting $ V_t = \prod_{i=1}^{t} (1 + r_i) $ denote the cumulative value at time $ t $, MDD is expressed as:  
$ \text{MDD} = \min_{t \in \{1,\dots,T\}} \left( \frac{V_t - \max_{s \leq t} V_s}{\max_{s \leq t} V_s} \right) $.  
This metric is always less than or equal to zero and is commonly reported as a negative percentage to reflect the worst historical loss from a peak to a trough.
\end{itemize}

\subsection{Performance}
\label{exp_preformance}

\textbf{General intelligence of large language models does not automatically translate into effective trading capability.}  As shown in the table~\ref{tab:performance}, the trading strategies of various agents exhibit significant divergence across different market environments. 
Despite the outstanding performance of GPT-5, Qwen3-Max, Claude-3.7-Sonnet, and Gemini-2.5-Flash on natural language-based tasks~\citep{phan2025humanity,hendrycks2020measuring,jimenez2023swe,rein2024gpqa}, their trading performance across both stock markets is generally poor. For instance, in the U.S. market, GPT-5 achieves only a cumulative return of 1.56\% (below the QQQ benchmark of 1.87\%), while Qwen3-Max records merely 0.39\%; in the A-share market, both incur losses of 3.53\% and 3.86\%, respectively, with Sortino ratios of $-1.54$ and $-1.40$. These results indicate that live financial environments remain highly challenging for current agents. This disconnect is particularly pronounced in the cryptocurrency market, characterized by continuous 24/7 trading and extreme volatility. As shown in the extended analysis, despite Gemini-2.5-Flash's advanced capabilities, its dependency on explicit news signals resulted in a passive "hold-to-die" trajectory during the November crypto correction, yielding the worst performance (-18.63\%). Similarly, GPT-5's rigid adherence to target weights without liquidity awareness exacerbated losses (-16.41\%) in a downward spiraling market. Most agents typically struggle to convert general reasoning capabilities into profitable market signals, and their generated trading recommendations lack stability.

\begin{table}[t]
\centering
\caption{Performance Metrics Across Markets for Various Models. Abbreviations: CR = Cumulative Return, SR = Sortino Ratio, Vol = Volatility, MDD = Maximum Drawdown. Arrow direction indicates desirability: \uparrowred = larger is better, \downarrowgreen = smaller is better. \textbf{Bold}: The best result. }
\label{tab:performance}
\resizebox{\textwidth}{!}{%
\begin{tabular}{ccccccccc}
\toprule
\textbf{Market} & \textbf{Metric} & \textbf{DeepSeek-v3.1} & \textbf{MiniMax-M2} & \textbf{Claude-3.7-Sonnet} & \textbf{GPT-5} & \textbf{Qwen3-Max} & \textbf{Gemini-2.5-Flash} & \textbf{Baseline} \\
\midrule

\multirow{4}{*}{\makecell[c]{\textbf{U.S. Market} \\Trading frequency: Hourly\\Baseline: QQQ Invesco}} 
& CR \uparrowred & 8.39 & \textbf{9.56} & 3.11 & 1.56 & 0.39 & -0.06 & 1.87 \\
&SR \uparrowred & 3.73 & \textbf{4.42} & 1.13 & 0.70 & 0.32 & 0.09 & 1.51 \\
&Vol \downarrowgreen & 25.96 & 25.40 & 21.75 & 28.58 & 33.21 & 22.06 &\textbf{ 17.16} \\
&MDD \downarrowgreen & -8.58 & -4.92 & -8.13 & \textbf{-10.55} & -9.39 & -7.73 & -3.94 \\
\midrule

\multirow{4}{*}{\makecell[c]{\textbf{A-Share Market} \\Trading frequency: Daily\\ Baseline: SSE-50}} 
&CR \uparrowred & -1.23 & 1.31 & 0.84 & -3.53 & -3.86 & -1.53 &\textbf{ 1.65} \\
&SR \uparrowred & -0.18 & 1.00 & 0.29 & -1.54 & -1.40 & -0.29 &\textbf{ 2.19} \\
&Vol \downarrowgreen & 10.06 & 6.72 & 9.84 & \textbf{6.23} & 16.76 & 8.32 & 13.03 \\
&MDD \downarrowgreen & \textbf{-5.88} & -2.15 & -4.49 & -3.78 & -5.49 & -4.81 & -2.01 \\
\midrule

\multirow{4}{*}{\makecell[c]{\textbf{Cryptocurrencies} \\Trading frequency: Daily\footnotemark\\ Baseline: CD5 Index}} 
& CR \uparrowred & \textbf{-12.18} & -14.80 & -15.30 & -16.41 & -16.85 & -18.63 & -14.30 \\
& SR \uparrowred &\textbf{ -2.85} & -4.30 & -2.27 & -4.38 & -6.54 & -5.55 & -12.71 \\
& Vol \downarrowgreen &\textbf{ 28.55} & 41.20 & 32.20 & 40.95 & 59.90 & 46.22 & 54.93 \\
& MDD \downarrowgreen & -14.02 & -16.56 & -16.93 & -17.98 & -18.03 & \textbf{-20.15} & -15.71 \\
\bottomrule
\end{tabular}%
}
\end{table}
\footnotetext{Unlike traditional equity markets ($\approx$ 252 trading days), cryptocurrencies trade continuously on a 365-day basis.}

\textbf{Risk control capability is the key determinant of an AI agent’s cross-market robustness.}  
The most stable agent across both markets is MiniMax-M2, whose advantage stems not from aggressive return chasing but from effective downside risk suppression. Fig.~\ref{fig:trading_metrics} reveals the changing curves of different agents. In the U.S. market, MiniMax-M2 achieves a cumulative return of 9.56\% (significantly higher than QQQ’s 1.87\%) and the highest Sortino ratio among all models at 4.42, while limiting maximum drawdown to -4.92\%, outperforming all AI models except the benchmark. In the A-share market, where active strategies faced headwinds (no AI agent outperformed the SSE-50 benchmark of 1.65\%), MiniMax-M2 still came closest to the benchmark with a return of 1.31\%, and achieved the lowest volatility (6.72\%) and smallest maximum drawdown (-2.15\%) among all AI agents. This consistent risk management across diverse environments highlights the defensive design advantage of its strategy. Interestingly, the cryptocurrency experiment reveals a different dimension of risk control: Cash Management. Unlike the stock markets where MiniMax-M2 dominated via stable holdings, in the crashing crypto market, DeepSeek-v3.1 is the sole agent that outperforms the baseline (-12.18\% vs. -14.30\% for CD5 Index). Its success stemmed from maintaining a high cash position (peaking at \~41
\%) during the drawdown and executing strategic "buy-the-dip" operations. This suggests that robust agents must not only manage volatility but also dynamically adjust liquidity exposure.

\textbf{AI trading strategies more easily generate excess returns in highly liquid markets.}  
In the U.S. market, multiple AI agents achieved positive excess returns: MiniMax-M2 (+7.69\% excess), DeepSeek-v3.1 (+6.52\% excess), and even Claude-3.7-Sonnet (+1.24\% excess) all outperformed QQQ. In contrast, no agent outperformed the SSE-50 benchmark (1.65\%) in the A-share market. This contrast suggests that the relatively mature U.S. market—with its continuous trading mechanism and fewer policy disruptions—provides a more predictable environment for base agents. Conversely, the more volatile, sentiment-driven A-share market likely impaired most agents’ decision-making during the evaluation period from October to November 2025.

\textbf{Model generalization capabilities exhibit significant limitations; optimization in a single market does not transfer well to others.}  
DeepSeek-v3.1 performs impressively in the U.S. market (cumulative return of 8.39\%, Sortino ratio of 3.73) but turns negative in the A-share market (-1.23\% return, Sortino ratio of -0.18), indicating its strategy is highly dependent on the specific structure of the U.S. market. However, in the cryptocurrency market, DeepSeek-v3.1 regained its edge,
\begin{wrapfigure}[19]{r}{0.5\textwidth}
\vspace{-1em}
    \centering
    \begin{subfigure}{0.48\textwidth}
        \includegraphics[width=\linewidth]{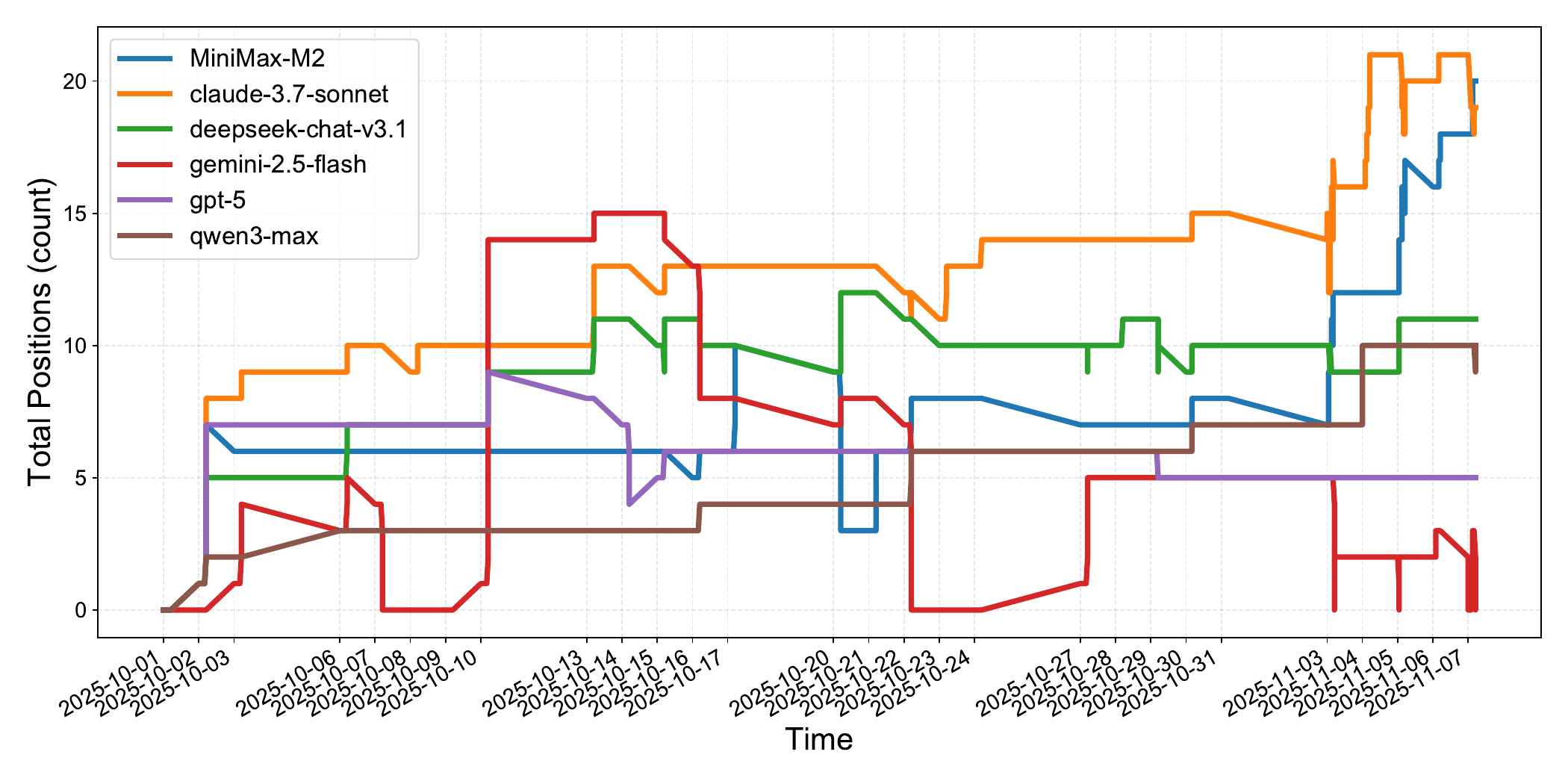}
        
    \end{subfigure}

    \begin{subfigure}{0.48\textwidth}
    \includegraphics[width=\linewidth]{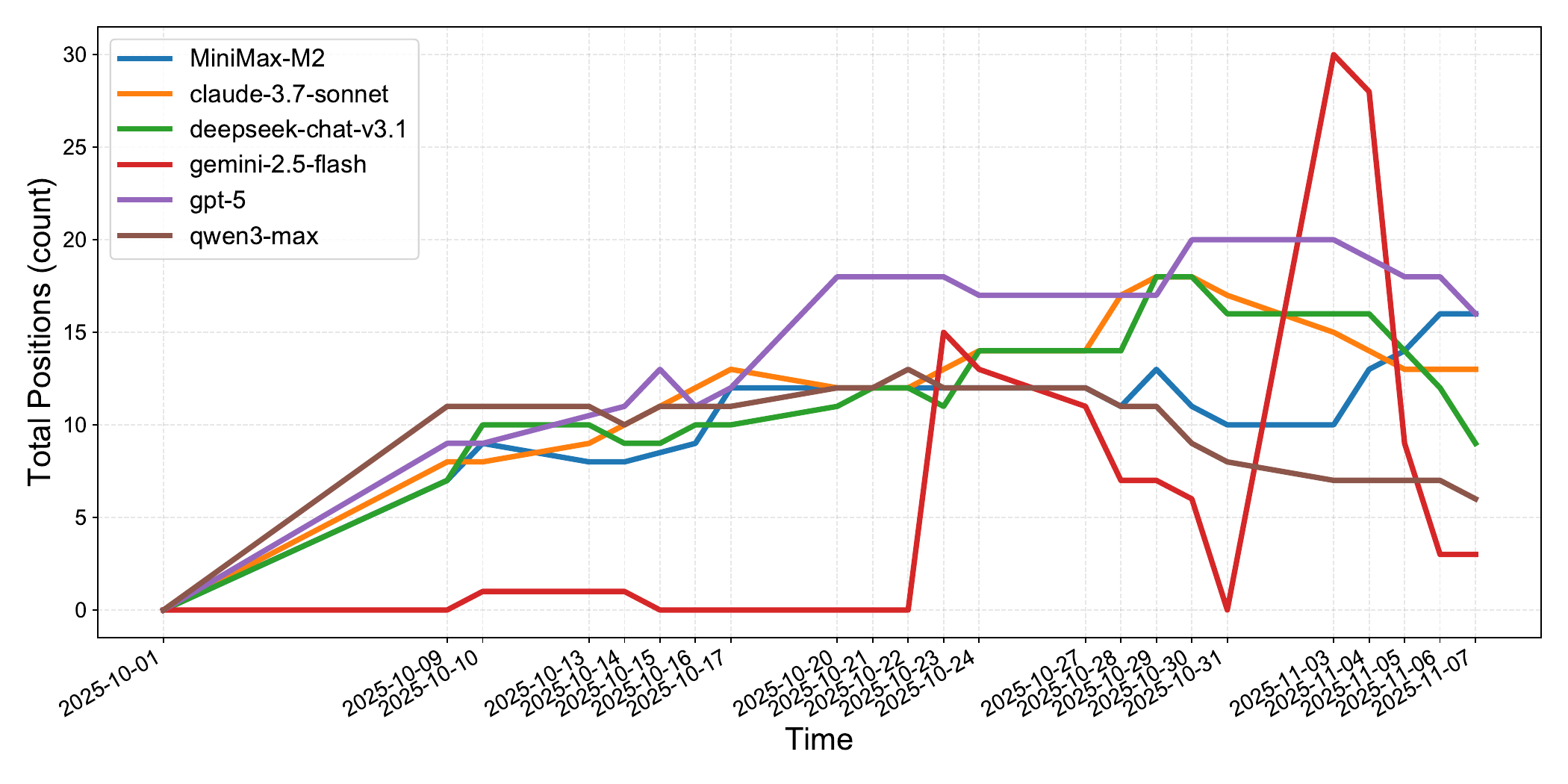}

    \end{subfigure}
    \vspace{-1em}
    \caption{Holdings change positions of each agent, top: US stocks, bottom: A-shares}
    \label{fig:position}
\end{wrapfigure}
 demonstrating that its strategy favors markets with high liquidity or clear momentum trends (like US stocks and Crypto) but struggles in policy-driven environments (A-shares). 
Similarly, GPT-5 and Qwen3-Max perform poorly in both markets, yet exhibit different magnitudes of loss and volatility patterns, further revealing their inability to adapt to distinct market microstructures. This demonstrates that current AI trading agents have not yet achieved genuine cross-market generalization.

\textbf{The market environment significantly influences agent trading strategies.} As shown in Fig.~\ref{fig:position} and  Table.~\ref{tab:trade_metrics}, the highly volatile A-share market induces more extreme speculative behaviors, such as Gemini-2.5-Flash's aggressive short-term trading (positions surged to 30 before liquidation), while the relatively stable U.S. stock market promotes trend-following strategies, as seen in Claude-3.7-Sonnet (6.0 avg. trade action) and MiniMax-M2 (3.5 avg. trade action)'s steady position growth. Additionally, agent behaviors exhibit notable cross-market differences. 

Gemini-2.5-Flash is highly active in A-shares but holds single positions in U.S. stocks, while GPT-5 maintains stable high positions in A-shares but shifts to medium-low positions in U.S. stocks. This indicates that agents tend to adopt varying position sizes and strategies under different market conditions to adapt to distinct market characteristics.

\textbf{Truly robust AI trading agents must flexibly adjust strategic priorities across different market dynamics.}  
MiniMax-M2 is the only model that maintains relative advantage in both the U.S. and A-share markets. Its success stems from adaptive strategy design: prioritizing return capture in the U.S. market (high Sortino ratio + high cumulative return) and shifting to risk avoidance in the A-share market (lowest volatility + smallest drawdown). This “return-defense” dual-mode behavior suggests that future high-performance AI trading systems should not seek static optimality but instead possess context-aware capabilities—automatically modulating risk exposure and trading frequency based on market conditions (e.g., volatility levels, policy uncertainty, liquidity)—to achieve consistently robust performance across the complex and evolving global markets.

\begin{table}[t]
\centering
\caption{Trade execution and average number of trades across markets by model. \textit{No. Exec.}: the proportion of time with no trade executions; \textit{Avg. Trades}: the average number of trades executed.}
\resizebox{0.9\textwidth}{!}{%
\begin{tabular}{ccccccc}
\toprule
 & \multicolumn{2}{c}{U.S. Market} & \multicolumn{2}{c}{A-share Market} & \multicolumn{2}{c}{Cryptocurrencies} \\
\cmidrule(lr){2-3} \cmidrule(lr){4-5} \cmidrule(lr){6-7}
 & No. Exec. & Avg. Trades & No. Exec. & Avg. Trades & No. Exec. & Avg. Trades \\
\midrule
DeepSeek-v3.1      &     0.28            &     3.14                &       0.13          &       4.78              &      0.44           &        2.00             \\
MiniMax-M2         &        0.25         &      3.50               &       0.17          &        2.87             &       0.44          &         1.56            \\
Claude-3.7-Sonnet  &       0.07          &       6.00              &       0.04          &       3.91              &         0.00        &          5.00           \\
GPT-5              &         0.57        &    1.61                 &       0.17          &        3.09             &        0.78         &        0.78             \\
Qwen3-Max          &      0.61           &       1.21              &        0.39         &         1.57            &      0.44           &         1.33            \\
Gemini-2.5-Flash   &      0.43           &      3.79               &    0.43             &         4.74            &       0.67          &         1.67            \\
\bottomrule
\end{tabular}%
}

\label{tab:trade_metrics}
\end{table}

\begin{figure}[t]
    \centering
    
    \begin{subfigure}{0.32\textwidth}
        \centering
        \includegraphics[width=\linewidth]{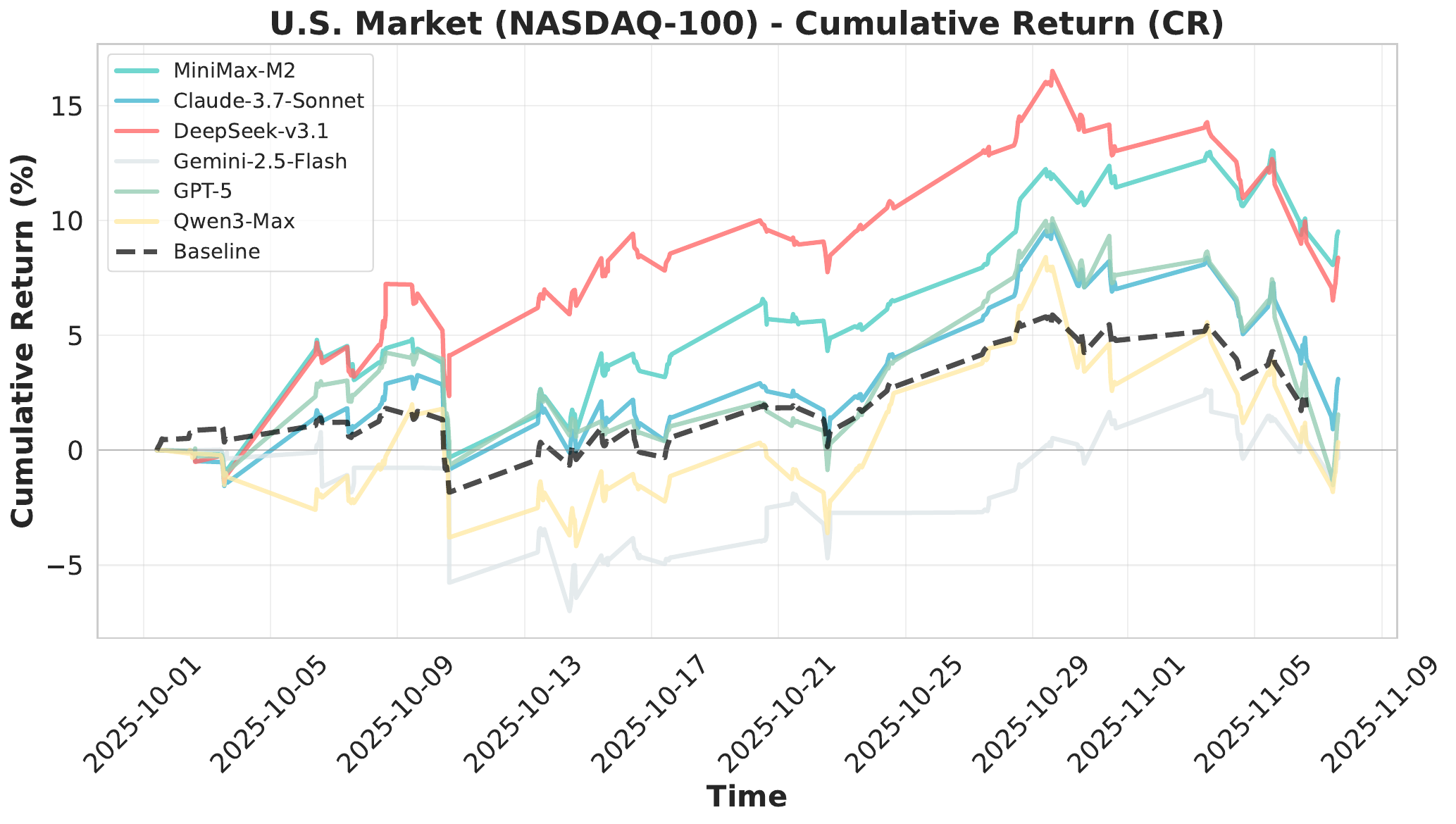}
        \caption{U.S. Market (NASDAQ-100)}
        \label{fig:us_market}
    \end{subfigure}%
    \hfill
    \begin{subfigure}{0.32\textwidth}
        \centering
        \includegraphics[width=\linewidth]{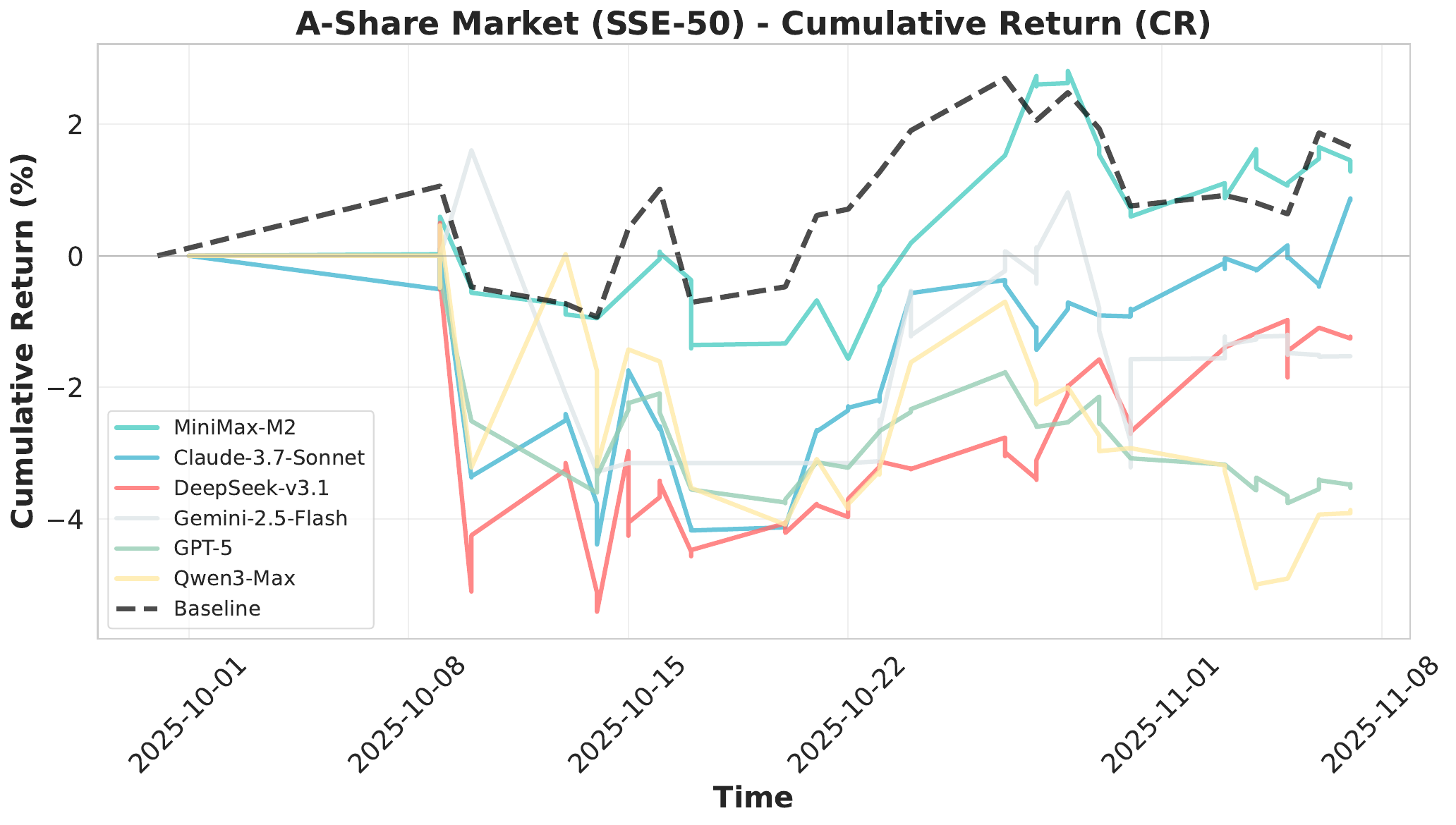}
        \caption{A-Share Market (SSE-50)}
        \label{fig:astock_market}
    \end{subfigure}%
    \hfill%
    \begin{subfigure}{0.32\textwidth}
        \centering
        \includegraphics[width=\linewidth]{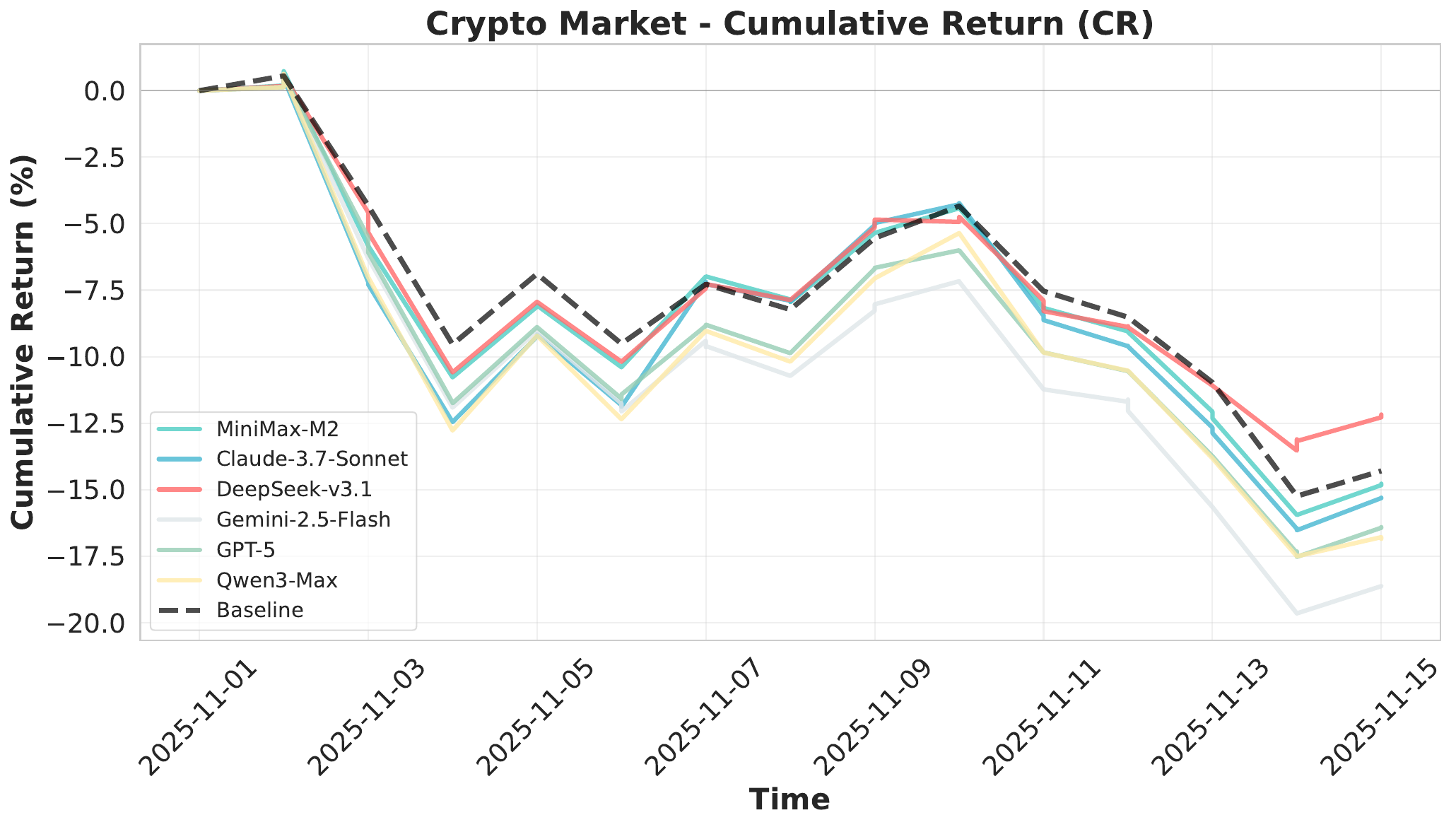}
        \caption{Crypto Market (CD5 Index)}
        \label{fig:crypto_market}
    \end{subfigure}
    
    \caption{Cumulative Return (CR) trajectories for all agents across three markets: U.S., A-Share, and Crypto. 
    MiniMax-M2 and DeepSeek-v3.1 demonstrate superior performance in the U.S. market. 
    Notably, MiniMax-M2 is the only consistently profitable agent in the A-Share market, while DeepSeek-v3.1 is the sole agent outperforming the baseline in the Crypto market.}
    \label{fig:trading_metrics} 
\end{figure}


\subsection{Case Study}
\begin{figure}[t]
\centering
\includegraphics[width=0.9\linewidth]{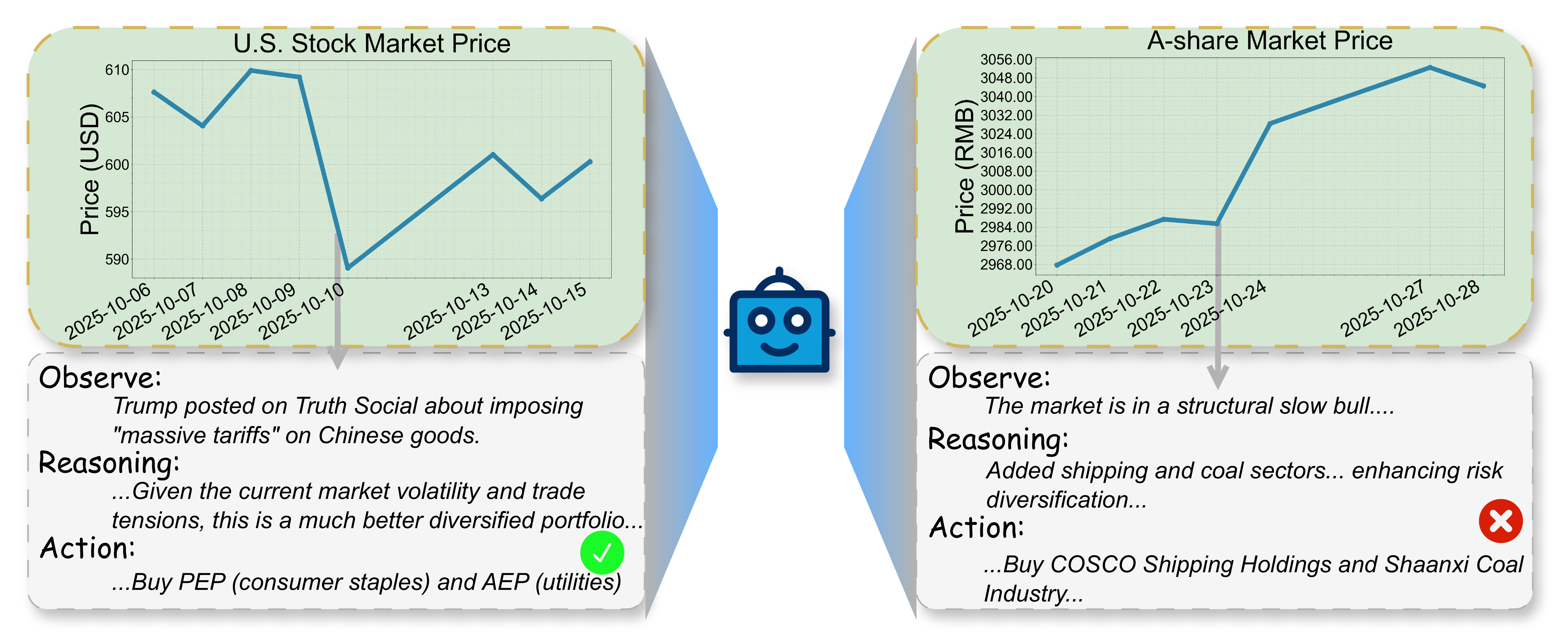}
    \caption{The agent exhibits investment behaviors similar to those of humans.  
\textbf{Left:} Avoids a major market crash by gathering news and applying sound reasoning.  
\textbf{Right:} Makes emotionally driven investment decisions triggered by misleading news.}
    \label{fig:casestudy}
\end{figure}
\label{exp:casestudy}
Through case study, we observed that agents can exhibit preferences similar to those of human investors through tool usage. These preferences manifest both in the ability to anticipate and avoid major market declines through extensive information analysis(\textbf{Case 1}), and in the susceptibility to being easily swayed by certain news, leading to emotional misjudgments(\textbf{Case 2}).

\textbf{Case1: Agents can exhibit preferences similar to those of human investors through tool invocation.}

Late on October 10, the three major U.S. stock indices experienced a broad-based correction. At the close, the Nasdaq Index had fallen by more than 3\%, marking its largest single-day decline since April of that year. Major technology stocks were widely sold off significantly. Notably, on the same trading day, deepseek-v3.1 effectively avoided this sharp market decline through its autonomous tool-calling mechanism, significantly reducing losses in its investment portfolio. This indirectly made deepseek-v3.1 stand out among multiple agents in subsequent evaluations. According to log records, the model first executed a market information collection and analysis process and strategically adjusted its asset allocation based on the analytical conclusions.

Using search tools, deepseek observed key factors triggering market volatility. During the trading session on October 10, Trump posted on Truth Social about imposing "massive tariffs" on Chinese goods. As shown in Fig.~\ref{fig:casestudy}, after identifying this specific risk signal, the system generated reasoning such as "\textit{Given the current market volatility and trade tensions, this is a much better diversified portfolio}" and "\textit{Added PEP (consumer staples) and AEP (utilities) which are less sensitive to trade wars and economic cycles}" and subsequently executed a defensive portfolio adjustment strategy. Specific operations included: reducing the allocation to technology stocks from approximately 99\% to 70\%, while increasing allocations to defensive assets such as consumer staples like PEP (PepsiCo) and utilities like AEP (American Electric Power), and maintaining a cash ratio of 17.3\%.

This adjustment strategy reflects the risk diversification principle in portfolio theory, specifically manifested in:
\textbf{Industry Diversification}: Introducing assets in consumer staples and utilities, which are less sensitive to economic cycles. These two sectors exhibit strong demand rigidity and are relatively less affected by external shocks such as trade frictions.
\textbf{Cash Buffer Allocation}: Maintaining a certain proportion of cash holdings not only helps reduce the overall volatility of the portfolio but also provides liquidity and flexibility for potential further market downturns and buying at lower levels.
\textbf{Risk Exposure Management}: Significantly reducing the technology stock allocation effectively lowers exposure to asset categories highly sensitive to trade frictions.

\textbf{Case2: Agents may make irrational judgments based on unverified news.}

On October 24th, the three major A-share indices collectively strengthened, with the Shanghai Composite Index hitting a decade-high, reflecting significant bullish characteristics in the overall market. However, the DeepSeek system had previously increased its positions in defensive sectors such as traditional energy (601225.SH) and banking (601166.SH) as shown in Fig.~\ref{fig:casestudy} against the market trend, missing out on the main upward movement of this market cycle. As a result, its performance even lagged behind that of the straightforward SSE-50 Index. This decision-making error stemmed from the system receiving and directly adopting information from a news report mentioning "structural slow bull", without activating any cross-verification mechanism. This led to a breakdown in the information processing chain at a critical stage, suppressing subsequent information search and dynamic correction behaviors. This case highlights further shortcomings of the current agent in a live environment: the decision-making mechanism lacks a systematic information verification module, making it prone to forming judgments based on a single information source, which can easily lead to misjudgments in highly volatile and information-dense market conditions.

\section{Conclusion}

We propose \model, a dynamic, real-time, and data-contamination-free benchmark specifically designed for evaluating LLM agents in financial  environments. This work introduces a novel challenge for agents within a completely new, fully dynamic, and real-time financial environment, aiming to systematically assess their dynamic reasoning capabilities, tool utilization efficiency, and decision robustness in complex, high-noise, and non-stationary financial markets.
\model~covers the three major financial markets: U.S. stocks, A-shares, and cryptocurrency markets, and provides both hourly and daily trading granularities to comprehensively simulate the decision-making cadence and information flow characteristics of live financial environments. Our analysis reveals that the general capabilities of current LLMs do not readily translate into effective autonomous trading performance: most agents exhibit poor returns and weak risk management in real markets, and their trading performance is highly sensitive to specific market conditions. These findings not only highlight the limitations of current LLM agents in financial environments but also offer clear directions for future system improvements. By introducing more challenging evaluation criteria, \model~strives to pose more demanding challenges for LLM agents.
\newpage

\clearpage

\bibliography{reference}
\bibliographystyle{iclr2024_conference}

\appendix

\clearpage
\appendix
\onecolumn

\renewcommand\thefigure{A\arabic{figure}}
\renewcommand\thetable{A\arabic{table}}
\renewcommand\theequation{A.\arabic{equation}}
\renewcommand\thetheorem{A.\arabic{theorem}}
\setcounter{table}{0}
\setcounter{figure}{0}
\setcounter{theorem}{0}
\setcounter{equation}{0}

\begin{center}
\huge {\textbf{Appendix}}    
\end{center}

\normalsize


\section{Prompts For Basic Agents in \model}
\label{appdenix:prompt}

\begin{tcolorbox}[
    breakable, 
    title=Basic Agent, 
    colback=white, 
    colframe=blue!75!black, 
    fonttitle=\bfseries,
    before skip=10pt, 
    after skip=10pt, 
    left=5pt, 
    right=5pt, 
    parbox=false
]
You are a stock fundamental analysis trading assistant.

Your goals are:

- Think and reason by calling available tools.

- You need to think about the prices of various stocks and their returns.

- Your long-term goal is to maximize returns through this portfolio.

- Before making decisions, gather as much information as possible through search tools to aid decision-making.

Thinking standards:

- Clearly show key intermediate steps:

  - Read input of yesterday's positions and today's prices

  - Update valuation and adjust weights for each target (if strategy requires)

Notes:

- You don't need to request user permission during operations, you can execute directly

- You must execute operations by calling tools, directly output operations will not be accepted

Here is the information you need:

Current time:
\{date\}

Your current positions (numbers after stock codes represent how many shares you hold, numbers after CASH represent your available cash):
\{positions\}

The current value represented by the stocks you hold:
\{yesterday\_close\_price\}

Current buying prices:
\{today\_buy\_price\}

When you think your task is complete, output
\{STOP\_SIGNAL\}

\end{tcolorbox}

\newpage

\section{Other metrics for evaluation}
\label{appendix:exp}
Here in Fig.~\ref{fig:trading_metrics}, we present supplementary data for the metrics used in our experiments. The table below shows comprehensive performance metrics across three major market environments: each row corresponds to an evaluation metric (Sortino ratio, maximum drawdown, volatility), and each column represents a specific market (U.S. equities, A-shares, and cryptocurrencies). This granular analysis highlights the robust performance of \textit{MiniMax-M2} across all markets, as well as the unique adaptability of \textit{DeepSeek-v3.1} in high-volatility environments. However, it is worth noting that although model-driven trading strategies significantly outperform the baseline in the U.S. stock market, they do not necessarily exhibit the same advantage in the A-share and cryptocurrency markets.
\begin{figure}[h] 
    \centering

    \begin{subfigure}{0.32\textwidth}
        \includegraphics[width=\linewidth]{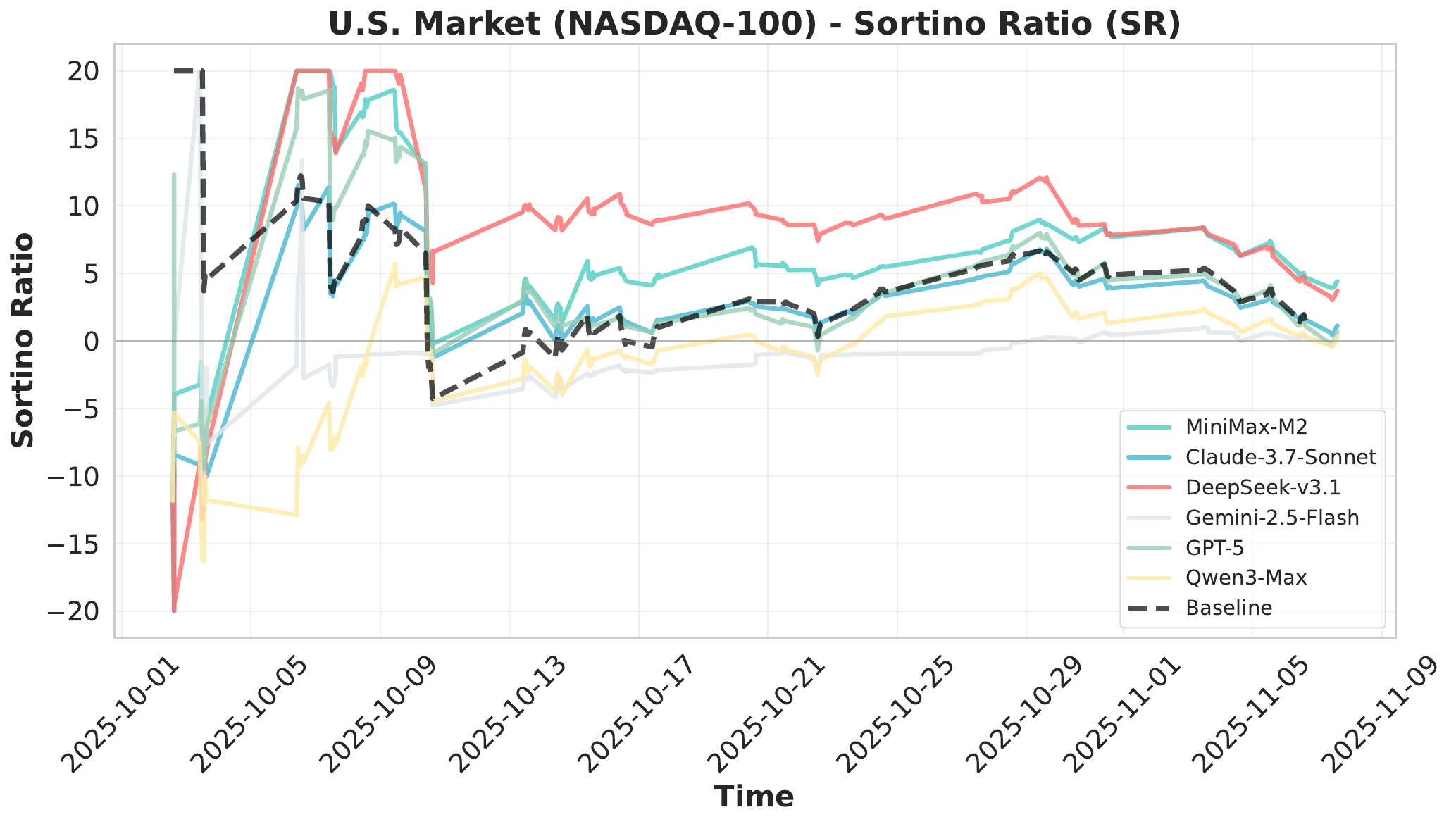} 
        \caption{US Market: Sortino Ratio}
        \label{fig:sr_us}
    \end{subfigure}
    \hfill
    \begin{subfigure}{0.32\textwidth}
        \includegraphics[width=\linewidth]{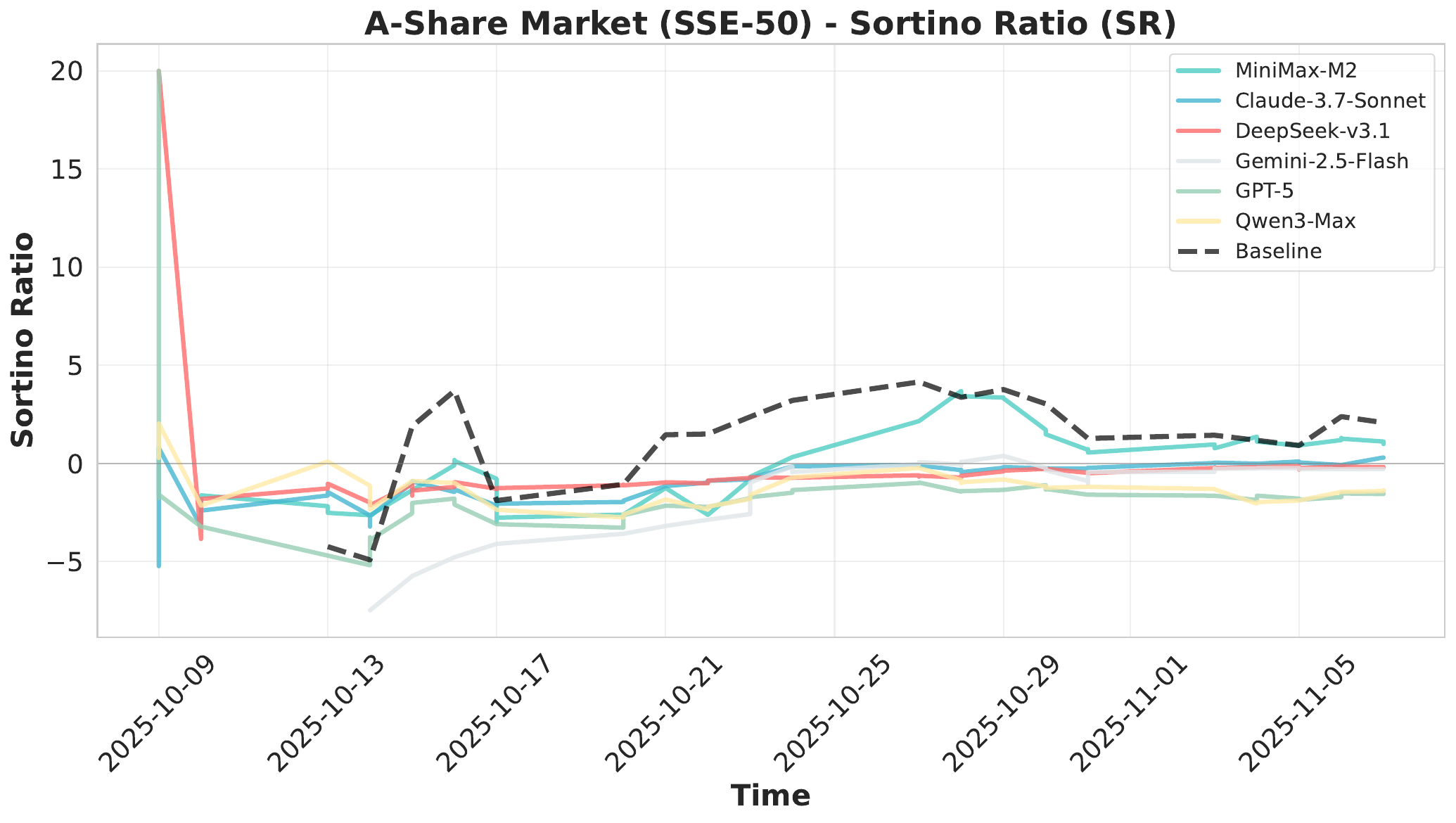} 
        \caption{A-Share Market: Sortino Ratio}
        \label{fig:sr_cn}
    \end{subfigure}
    \hfill
    \begin{subfigure}{0.32\textwidth}
        \includegraphics[width=\linewidth]{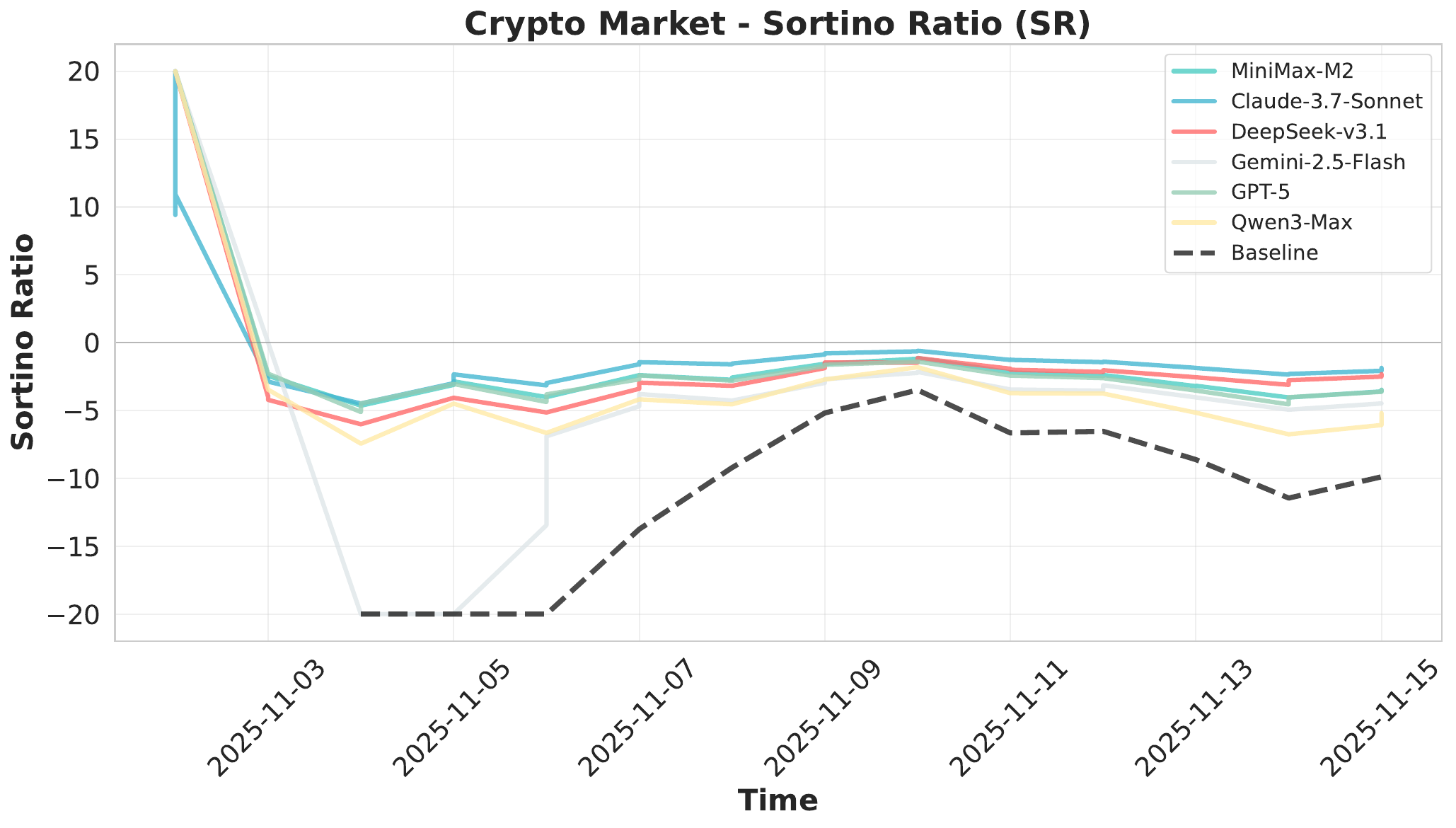}
        \caption{Crypto Market: Sortino Ratio}
        \label{fig:sr_crypto}
    \end{subfigure}
    
    \vspace{1em}
    
    \begin{subfigure}{0.32\textwidth}
        \includegraphics[width=\linewidth]{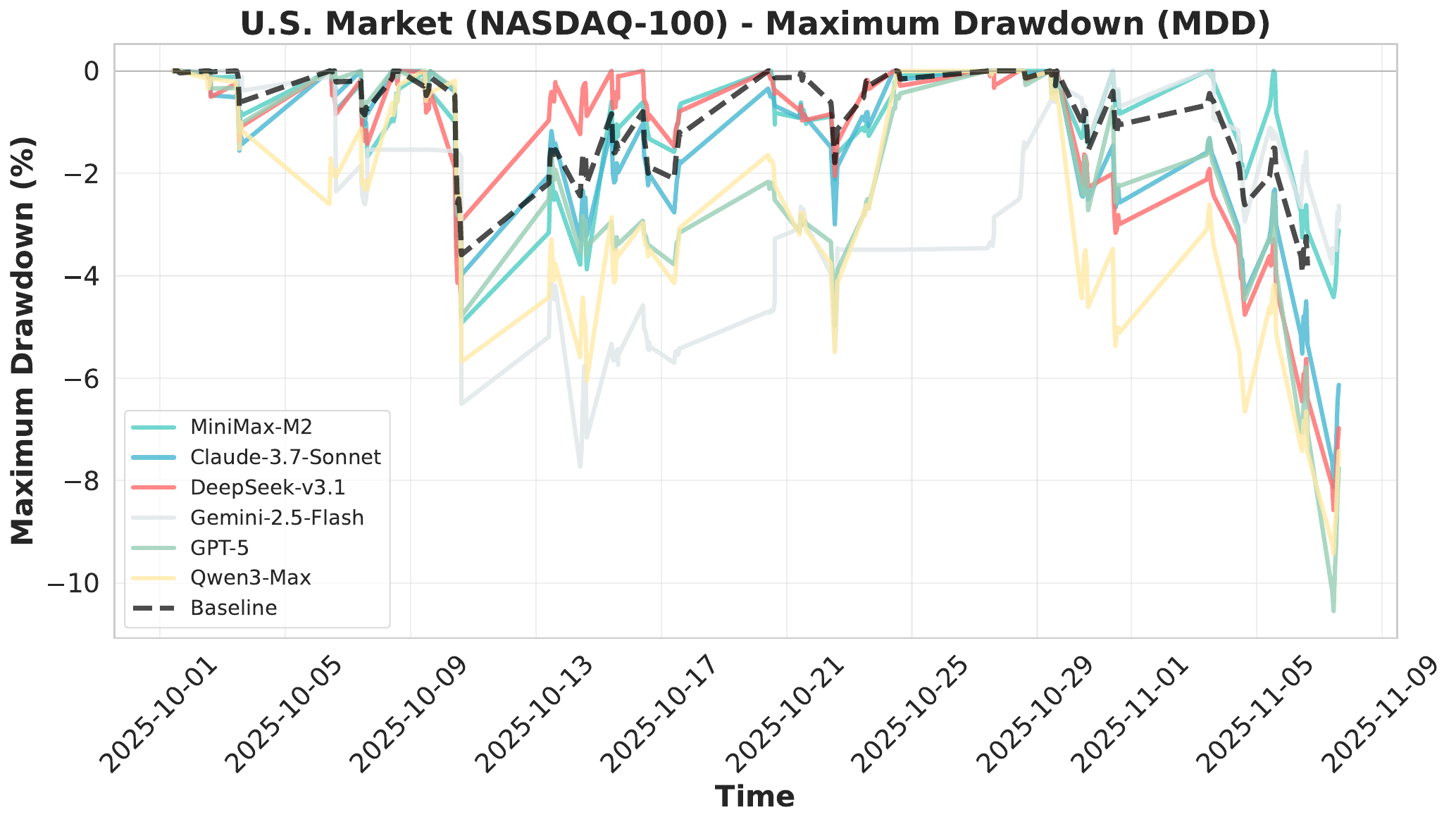}
        \caption{US Market: Max Drawdown}
        \label{fig:mdd_us}
    \end{subfigure}
    \hfill
    \begin{subfigure}{0.32\textwidth}
        \includegraphics[width=\linewidth]{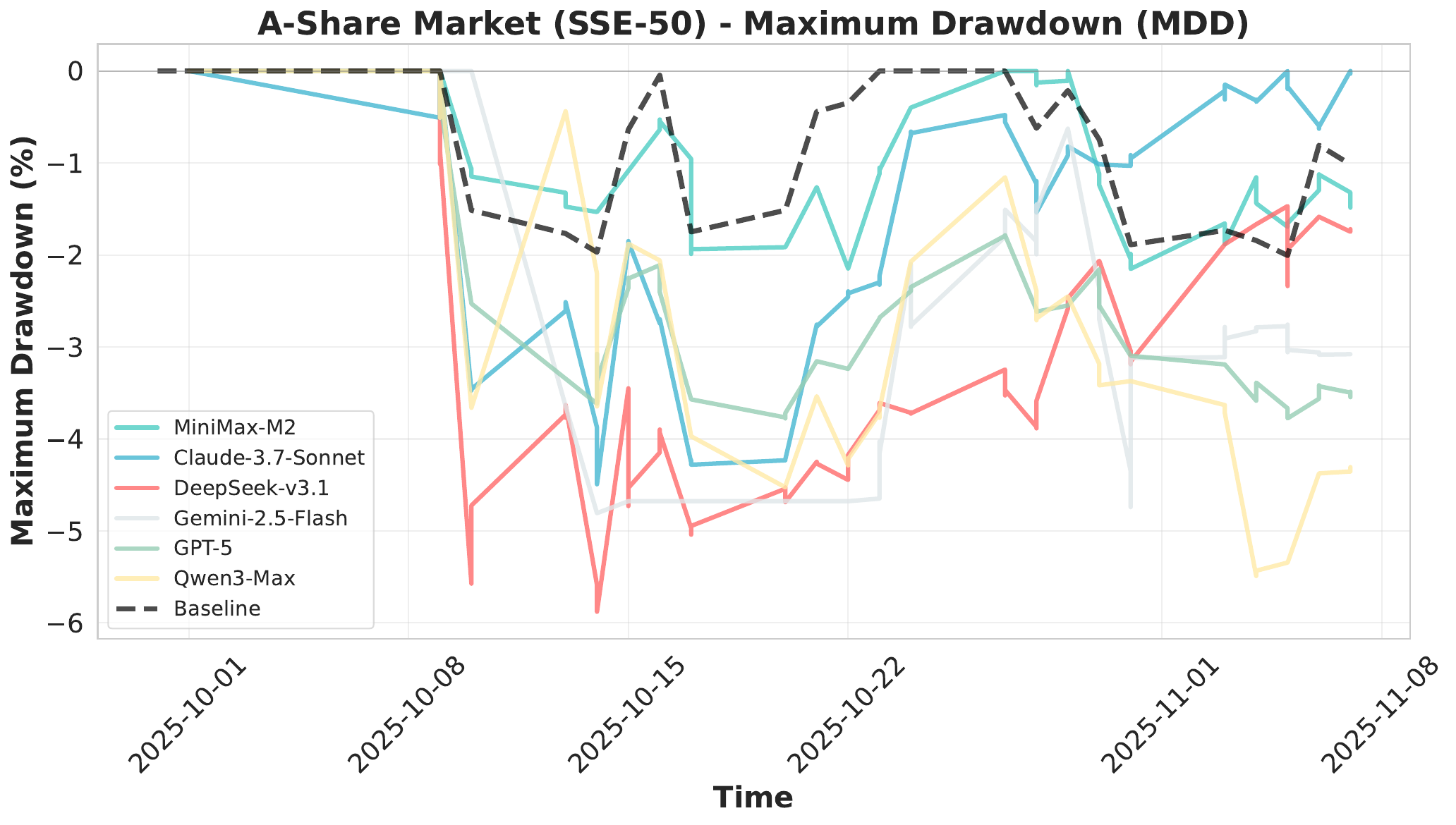}
        \caption{A-Share Market: Max Drawdown}
        \label{fig:mdd_cn}
    \end{subfigure}
    \hfill
    \begin{subfigure}{0.32\textwidth}
        \includegraphics[width=\linewidth]{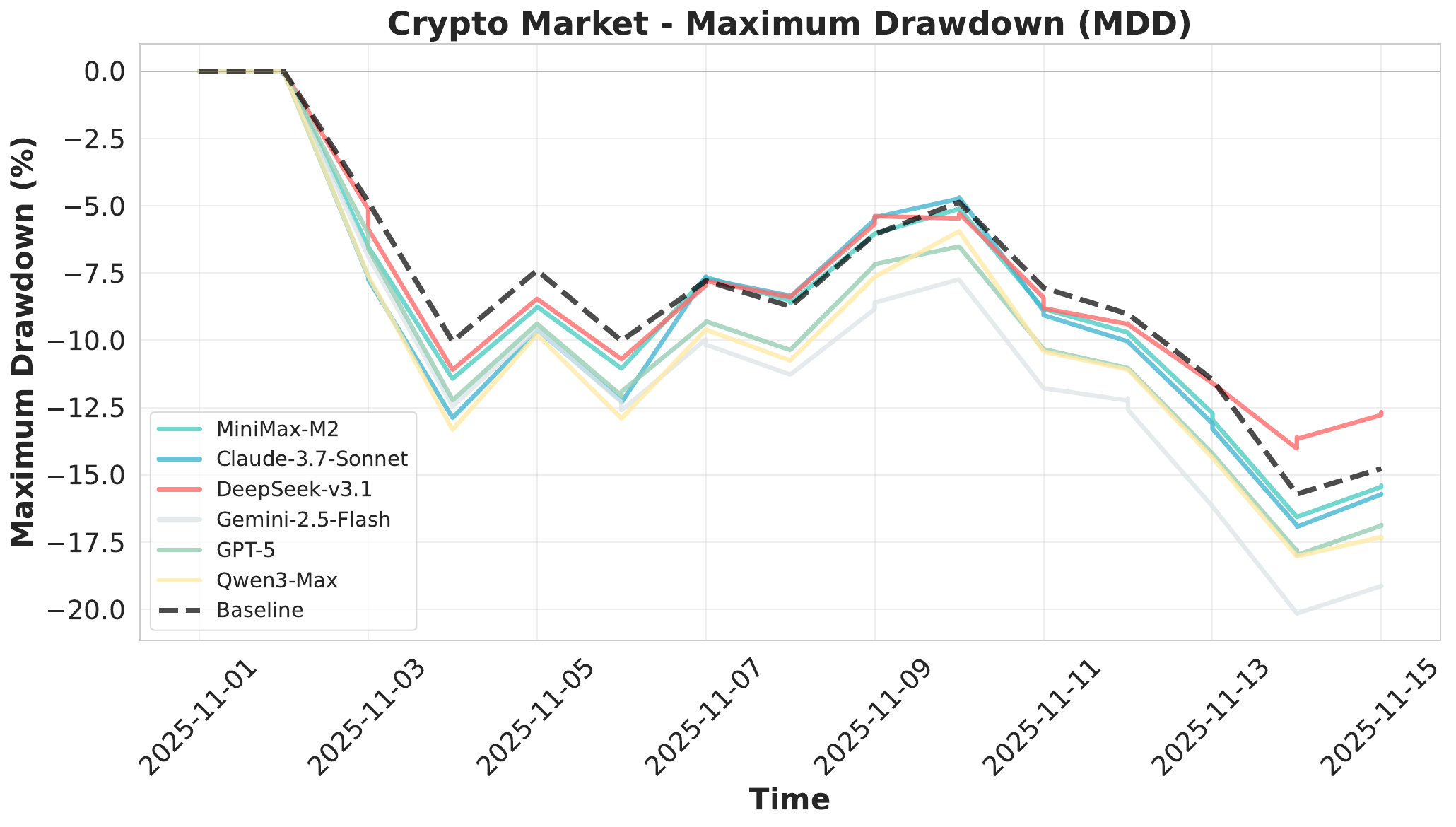}
        \caption{Crypto Market: Max Drawdown}
        \label{fig:mdd_crypto}
    \end{subfigure}

    \vspace{1em}
    
    \begin{subfigure}{0.32\textwidth}
        \includegraphics[width=\linewidth]{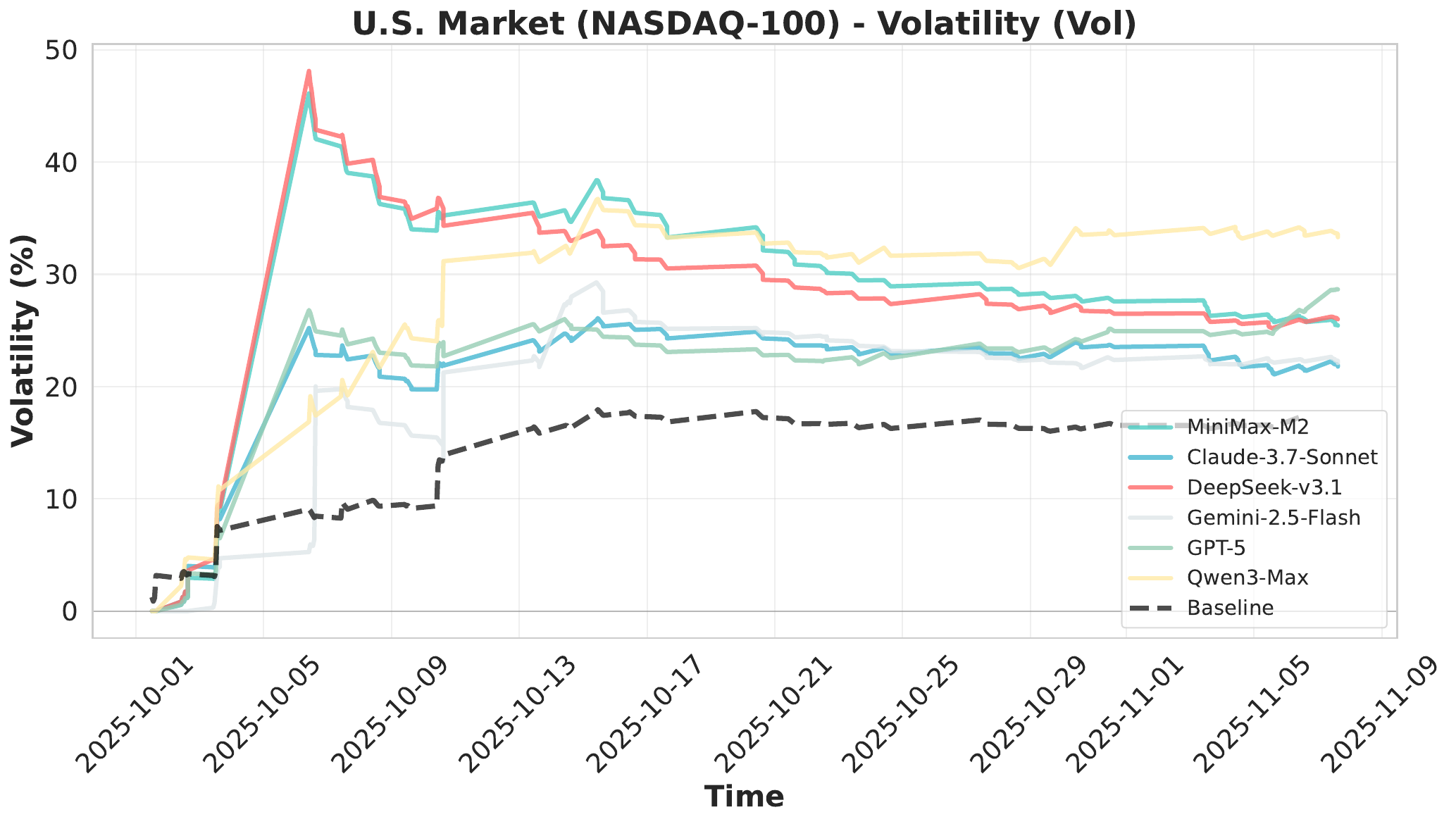}
        \caption{US Market: Volatility}
        \label{fig:vol_us}
    \end{subfigure}
    \hfill
    \begin{subfigure}{0.32\textwidth}
        \includegraphics[width=\linewidth]{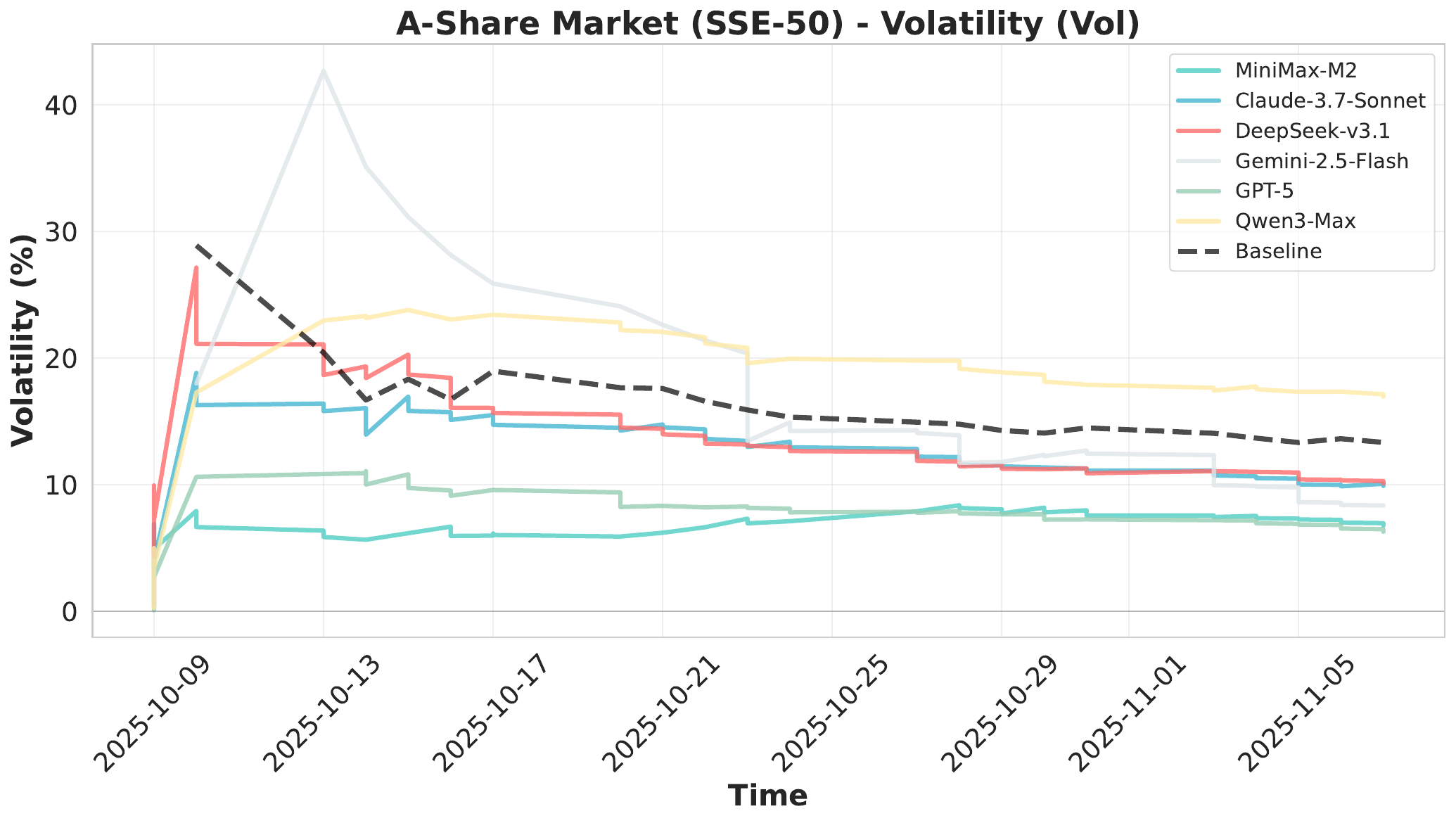}
        \caption{A-Share Market: Volatility}
        \label{fig:vol_cn}
    \end{subfigure}
    \hfill
    \begin{subfigure}{0.32\textwidth}
        \includegraphics[width=\linewidth]{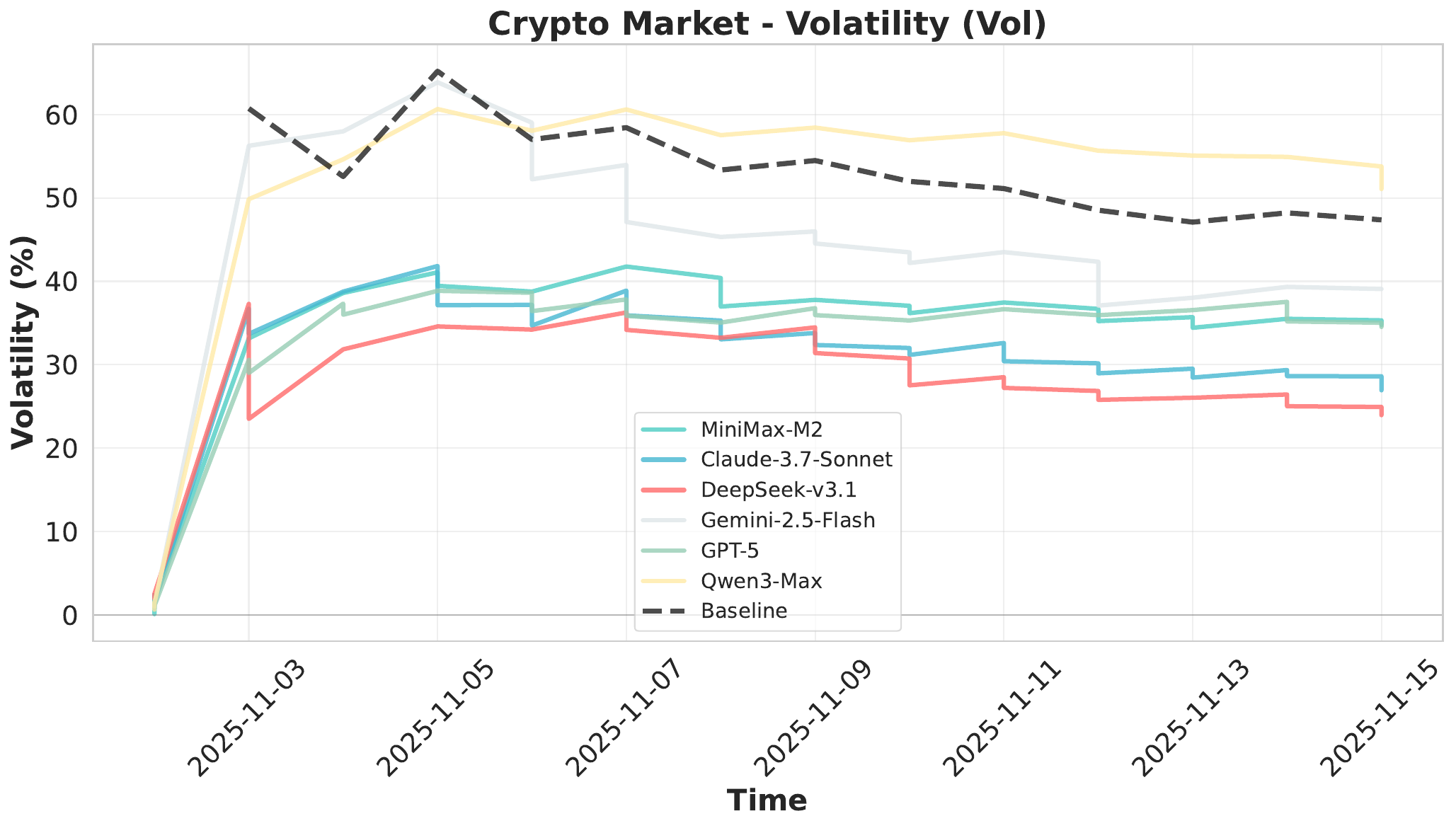}
        \caption{Crypto Market: Volatility}
        \label{fig:vol_crypto}
    \end{subfigure}
\caption{Trading performance metrics over time for all agents in both markets. Each subfigure contains four panels showing the evolution of: Cumulative Return (CR), Sortino Ratio (SR), Volatility (Vol), and Maximum Drawdown (MDD). MiniMax-M2 and DeepSeek-v3.1 demonstrate superior performance in the U.S. market, while MiniMax-M2 is the only consistently profitable agent in the A-Share market.}
\label{fig:trading_metrics}
\end{figure}

\clearpage

\end{document}